\documentclass[3p]{elsarticle}

\usepackage{amssymb}
\usepackage[utf8]{inputenc}

\newcounter{bla}

\journal{Computer Physics Communications}

\usepackage{booktabs}
\usepackage{amsfonts}
\usepackage{graphicx}
\usepackage{amsmath, amssymb}
\usepackage{listings}
\usepackage{bm}
\usepackage{floatrow}
\usepackage{hyperref}

\newfloatcommand{capbtabbox}{table}[][\FBwidth]

\newfloat{Listing}{t}{lop}

\newcommand{\bx}{\mathbf{x}}
\newcommand{\by}{\mathbf{y}}

\newcommand{\olp}{\bar{p}}

\newcommand{\olu}{\bar{u}}

\newcommand{\oltau}{\overline{\tau}}

\newcommand{\pdiff}[2]{\frac{\partial #1}{\partial #2}}
\newcommand{\diff}[2]{\frac{\mbox{d} #1}{\mbox{d} #2}}

\newcommand{\mean}[1]{\langle #1 \rangle}

\newcommand{\eps}{\epsilon}
\newcommand{\rey}{\mbox{Re}}
\newcommand{\ttt}[1]{\texttt{#1}}

\begin{document}


\begin{frontmatter}

\title{A Library for Wall-Modelled Large-Eddy Simulation Based on OpenFOAM Technology}

\author[focal1]{T.~Mukha\corref{cor1}}
\ead{timofey.mukha@it.uu.se}

\author[focal1]{S.~Rezaeiravesh}
\ead{saleh.rezaeiravesh@it.uu.se}

\author[focal1,focal2]{M.~Liefvendahl\corref{cor2}}
\ead{mattias.liefvendahl@foi.se}

\cortext[cor1]{Principal Corresponding Author}
\cortext[cor2]{Corresponding Author}
\address[focal1]{Uppsala University, Department of Information Technology, Box 337, SE-751 05 Uppsala, Sweden}
\address[focal2]{Swedish Defence Research Agency (FOI), SE-164 90 Stockholm, Sweden}


\begin{abstract}
This work presents a feature-rich open-source library for wall-modelled large-eddy simulation (WMLES), which is a turbulence modelling approach that reduces the computational cost of traditional (wall-resolved) LES by introducing special treatment of the inner region of turbulent boundary layers (TBLs).
The library is based on \ttt{OpenFOAM} and enhances the general-purpose LES solvers provided by this software with state-of-the-art wall modelling capability.
In particular, the included wall models belong to the class of wall-stress models that account for the under-resolved turbulent structures by predicting and enforcing the correct local value of the wall shear stress.
A review of this approach is given, followed by a detailed description of the library, discussing its functionality and extensible design.
The included wall-stress models are presented, based on both algebraic and ordinary differential equations.
To demonstrate the capabilities of the library, it was used for WMLES of turbulent channel flow and the flow over a backward-facing step (BFS).
For each flow, a systematic simulation campaign was performed, in order to find a combination of numerical schemes, grid resolution and wall model type that would yield a good predictive accuracy for both the mean velocity field in the outer layer of the TBLs and the mean wall shear stress.
The best result, $\approx 1\%$ error in the above quantities, was achieved for channel flow using a mildly dissipative second-order accurate scheme for the convective fluxes applied on an isotropic grid with $27\, 000$ cells per $\delta^3$-cube, where $\delta$ is the channel half-height.
In the case of flow over a BFS, this combination led to the best agreement with experimental data.
An algebraic model based on Spalding's law of the wall was found to perform well for both flows.
On the other hand, the tested more complicated models, which incorporate the pressure gradient in the wall shear stress prediction, led to less accurate results.
\end{abstract}

\end{frontmatter}


{\bf PROGRAM SUMMARY}

\begin{small}
\noindent
{\em Program Title: libWallModelledLES}                                          \\
{\em Licensing provisions:GPLv3}                                   \\
{\em Programming language: C++}                                   \\
{\em Nature of problem: Large-eddy simulation (LES) is a scale-resolving turbulence modelling approach providing a high level of predictive accuracy.
However, LES of high Reynolds number wall-bounded flows is prohibitively computationally expensive due to the need for resolving the inner region of turbulent boundary layers (TBLs)~[1].
This inhibits the application of LES to many industrially relevant flows~[2] and prompts for the development of novel modelling techniques that would modify the LES approach in a way that allows it to retain its accuracy (at least away from walls) yet significantly lowers its computational cost.}\\
{\em Solution method: Wall-modelled LES (WMLES) is an approach that is based on complementing LES with special near-wall modelling that allows to leave the inner layer of TBLs unresolved by the computational grid.
Many types of wall models have been proposed~[1,3], commonly tested within the framework of in-house research codes.
Here, an open-source library implementing several wall models is presented.
The library is based on \ttt{OpenFOAM}, which is currently the most widely-used general-purpose open-source software for computational fluid dynamics.
The developed library can be directly applied to both academic and industrial flow cases, leading to a wider adoption of wall modelling and better understanding of its strengths and limitations.}\\
   \\

\end{small}
\section{Introduction}
\label{sec:intro}

Large-eddy simulation (LES) is a scale-resolving turbulence modelling approach, which allows for accurate simulation of flow phenomena in cases where other methods, such as Reynolds-averaged Navier-Stokes (RANS), fail to provide a satisfactory result or when obtaining only the time-averaged values of the unknowns is insufficient.
One of the main obstacles to the wide adoption of LES for wall-bounded turbulent flows is the associated restrictive requirements on grid resolution near walls, which lead to the size of the grid being proportional to $\rey^{1.85}$~\cite{Choi2012, Rezaeiravesh2017, Liefvendahl2017b}, where $\rey$ is the Reynolds number.
The main reason for these resolution requirements is the structure of turbulence in the inner region of boundary layers, with the size of energetic eddies being on the viscous length scale, $\delta_\nu=\nu/u_\tau$, where $\nu$ is the kinematic viscosity, $u_\tau=\sqrt{\tau_w/\rho}$, with $\tau_w$ denoting the wall shear stress, and $\rho$ the fluid density,~\cite{Chapman1979,Sagaut2005}.
In contrast, in the outer part, the local turbulent boundary layer (TBL) thickness, $\delta$, is the relevant length scale.
The ratio of these two scales define the Reynolds number, $\rey_\tau = \delta/\delta_\nu$, indicating that an increasingly large fraction of the computational effort has to be put into resolving the inner region, as the Re-number grows.
This motivates introducing special modelling for the inner region, essentially aiming to resolve only the larger flow structures in the boundary layer, on the length scale $\delta$, and model the effect of smaller structures, of length scale $\delta_\nu$.
This general approach is referred to as wall-modelled LES (WMLES), for which the required number of grid points scales only linearly with $\rey$~\cite{Choi2012,Rezaeiravesh2016}, thus significantly extending the range of affordable Re-numbers.

There are different approaches to wall modelling, see~\cite{Sagaut2005,Piomelli2002, Piomelli2008, Larsson2016, Bose2018a} for reviews.
Here, a short overview of previous contributions is given to put the present work into context.
The most influential early study on WMLES appears to be that of Schumann,~\cite{Schumann1975}, where special boundary conditions were introduced at the wall in order to prescribe the correct value of the local wall shear stress.
Later, the way of computing this value and how it is enforced has been further developed, but the basic idea of accounting for the dynamics of the inner layer by prescribing the correct value of the filtered shear stress at the wall is still at the heart of a broad class of wall models, generally referred to as wall-stress models.

Schumann assumed that the mean value of the wall shear stress is known a priori.
In~\cite{Grotzbach1987}, the need for this restrictive assumption was removed by instead assuming that the mean velocity in the first off-the-wall grid point adheres to the log-law, allowing to compute the mean wall shear stress in the course of the simulation.
Later on, in~\cite{Cabot1995} and~\cite{Balaras1996}, more involved wall-stress models, based on both ordinary and partial differential equations (ODEs and PDEs), were developed. 
The premise was that such models would perform better in non-equilibrium flows, such as flows with separation.
This was examined by Wang and Moin \cite{Wang2002}, who applied several types of wall-stress models to a flow separating from the trailing edge of an airfoil, with a PDE-based model giving the best results in the recirculation region.
Further developments of PDE-based models, with focus on dynamic mechanisms for computing model parameters, were introduced in~\cite{Kawai2013, Park2014}.
The downsides of these models are the associated computational costs, as compared to simpler approaches, and also the difficulty of implementing them in solvers suited for unstructured meshes and complex domain geometries. 
Specifically, the latter implementational aspect has been recently addressed in~\cite{Park2016}. 
Most importantly, there is no consensus regarding whether PDE-based models are necessarily more accurate than less complicated approaches,~\cite{Larsson2016}.
The development of a wall-stress model based on ODEs or algebraic equations capable of simulation of non-equilibrium flows is a matter of on-going research.

Outside of wall-stress modelling, so-called hybrid LES/RANS approaches, such as Detached Eddy Simulation (DES)~\cite{Spalart1997} and others, provide alternatives.
Here, the part of the computational domain occupied by TBLs is separated (explicitly or implicitly) into a region where RANS equations are solved, and in the remainder of the domain, LES modelling is applied.
By instead constricting the RANS region to only include the inner region of the TBLs, the hybrid approaches can be adopted for WMLES, see e.g.~\cite{Nikitin2000}.
Yet another wall modelling approach, based on introducing a partial slip boundary condition for velocity, has been presented in~\cite{Bose2014}.

Independently of the employed wall modelling approach, a persistent issue has been the presence of a vertical shift in the obtained inner-scaled mean streamwise velocity profile, as compared to what is predicted by theory and direct numerical simulation (DNS). 
Mitigating this error, referred to as the log-layer mismatch (LLM), has been the focus of a significant number of studies.
Pertaining to wall-stress modelling, the following can be highlighted.
In~\cite{Kawai2012, Lee2013, Frere2017}, providing wall model input from a point located further from the wall is suggested.
Modifying the subgrid scale (SGS) viscosity close to the wall is recommended in~\cite{Wu2013}.
Recently in~\cite{Yang2017}, temporal filtering of the wall model input is proposed as a remedy instead.
All of these solutions are shown to be successful at removing the LLM, at least in conjunction with the other simulation parameters and numerical methods used in the respective studies.
However, consensus regarding the best approach has not yet been reached.

In the majority of the studies discussed above, the WMLES was conducted for flows with a relatively simple domain geometry and/or using in-house codes.
To the authors' best knowledge, no general-purpose computational fluid dynamics (CFD) code with advanced wall-stress modelling capabilities, is available under an open-source licence.
The main goal of this work is to present a newly-developed library for wall-stress modelling, based on \ttt{OpenFOAM} technology\footnote{The library is made available at \ttt{https://bitbucket.org/lesituu/libwallmodelledles}.}.
Originally developed as an in-house research code~\cite{Weller1998}, \ttt{OpenFOAM} is currently a publicly available general-purpose CFD software suite, enjoying a large user base and an active community.
Enhancing it with state-of-the-art WMLES capabilities can pave the way for more extensive validation of WMLES techniques and ultimately their wider adoption.
An early version of the presented library has been introduced in~\cite{Mukha2017a}.
Since then, its functionality has been significantly extended and several works on WMLES employing the library have been published~\cite{Mukha2018a, Mukha2018b, Liefvendahl2018}.
In this article, a full description of the wall-stress modelling capabilities currently provided by this software is given.
This includes models based on laws of the wall and ODEs, which are  both  discussed in detail in Section~\ref{sec:wss}.
A description of the libraries design and how it simplifies implementation of new wall-stress modelling approaches is given in Section~\ref{sec:library}.
This is expected to facilitate testing of new WMLES developments on flows defined by complicated geometries and in the numerical setting typical of modern industrial CFD solvers.
Key information about how to set up an \ttt{OpenFOAM} case to use the library is also provided here.
The flexibility of the configuration with respect to the choice of all the parameters controlling wall modelling is stressed.
This includes the possibility to arbitrarily choose the distance to the sampling point. 
Recall that adjusting this parameter was proposed as a method for mitigating the LLM.

The developed code is here applied to WMLES of two canonical wall-bounded turbulent flows: fully-developed turbulent channel flow and the flow over a backward-facing step.
The goal of these simulations is three-fold. 
One is to demonstrate the capabilities of the library and the predictive accuracy of the wall modelling approaches that it provides.
The second is to examine the effect of other modelling choices, such as the density of the grid and the employed numerical schemes. 
As a result, a combination of modelling parameters that results in good accuracy for the mean velocity profile in the outer layer and the mean wall shear stress is obtained.
Finally, by making the \ttt{OpenFOAM} set-up files for these simulations (as well as their results) available online\footnote{DOI: \ttt{10.6084/m9.figshare.6790013}.}, it is intended to provide new users a good starting point for setting up their own simulations.

The structure of the paper is as follows.
The CFD methods available in \ttt{OpenFOAM} and used here for WMLES are discussed in Section~\ref{sec:cfd}.
Wall-stress modelling is described in Section~\ref{sec:wss}.
Further, in Section~\ref{sec:library}, the design and features of the newly-developed library are presented.
In Section~\ref{sec:channel} and~\ref{sec:bfs}, the results from WMLES of fully-developed turbulent channel flow and flow over a backward-facing step are considered.
Concluding remarks concerning both the model implementations and the predictive accuracy are given in Section~\ref{sec:conclusions}.

\section{Computational fluid dynamics methods}
\label{sec:cfd}

The governing equations for LES are derived by applying spatial filtering to the incompressible Navier-Stokes equations, see e.g. \cite{Sagaut2005}.
The filtered momentum and continuity equations are,

\begin{eqnarray}
  \pdiff{\olu_i}{t}+\pdiff{}{x_j}\left(\olu_i\olu_j \right) & = & -\frac{1}{\rho} \pdiff{\olp}{x_i}
  + \pdiff{\tau_{ij}}{x_j},\quad (i=1,2,3) \label{eq:les} \\
  \pdiff{\olu_j}{x_j} & = & 0.
\end{eqnarray}
Here, summation is applied for repeated indices, and the overbar is used to denote spatially filtered quantities, e.g.~for the velocity,
\begin{equation} \label{eq:filt}
\olu_i(\bx,t) = \int G(\bx,\by)u_i(\by,t)\,\mbox{d}\by,
\end{equation}
for a filter kernel $G$.
The stress tensor, $\tau_{ij}$, consists of viscous and subgrid-scale terms\footnote{Strictly, $\rho \tau_{ij}$ is the stress tensor. However, here the mass-specific quantity is referred to by the same name, as is common when incompressible flow is considered, see e.g.~\cite{Pope2000}.},
\begin{eqnarray*}
\tau_{ij} =2\nu \bar{s}_{ij} + \tau^{\rm{\rm{sgs}}}_{ij} \,,
\end{eqnarray*} 
where, $\tau^{\rm{sgs}}_{ij}=-(\overline{u_i u_j}-\olu_i \olu_j)$, and the filtered rate-of-strain tensor is defined by,
$$
\bar{s}_{ij}=\frac{1}{2}\left(\frac{\partial \olu_i}{\partial x_j} + \frac{\partial \olu_j}{\partial x_i} \right) \,.
$$

LES modelling consists of providing a computable expression for the SGS tensor. 
A major class of models developed for this purpose is based on the Boussinesq approximation, in which the deviatoric part of the SGS stress tensor is modelled analogously to the viscous stress tensor,
\[
\tau^{\rm{sgs}}_{ij} = 2\nu_{\rm{sgs}} \bar{s}_{ij} + \frac{2}{3}k_{\rm{sgs}}\delta_{ij} \,,
\]
where $k_{\rm{sgs}} = \tau^\text{sgs}_{kk}/2$ and $\nu_{\rm{sgs}}$ is the SGS viscosity.
Many models have been developed for computing the latter, see~\cite{Sagaut2005} for a review.
The WALE model~\cite{Nicoud1999a} is used in the simulations presented in Sections~\ref{sec:channel} and \ref{sec:bfs}.
This choice is motivated by previous studies, e.g.~\cite{Temmerman2003}, where the WALE model was shown to perform well in conjunction with the relatively coarse grids typical of WMLES.

In \texttt{OpenFOAM}, the above governing equations are solved with the finite volume method over a spatial domain discretised on a grid consisting of arbitrary polyhedral cells.
In the collocated finite volume method, the unknowns are represented at the cell centres and approximate with second-order accuracy the average value of the respective quantities across the volume of the cell.
That is, for a cell with volume~$V_c$ and center point~$\bx_p$,
the cell-centred velocity value $\olu_i(\bx_p, t)$ approximates
\begin{equation}
\label{eq:cell_average}
\frac{1}{V_c}\int_{V_c} u_i(\by,t)\,\mbox{d}\by.
\end{equation}
Note that this is exactly of the form of the right-hand-side of equation~(\ref{eq:filt}) for the filter kernel, $G(\bx_p, \by)=H_c(\by)/V_c$, where $H_c$ is the Heaviside function corresponding to the cell.
This is thus the natural connection between the LES filtering and the finite volume framework, which directly connects the computational grid to the filtering operation.
For more details on the application of the finite volume method in computational fluid dynamics, see \cite{Ferziger2002}, and for formulation in connection with WMLES, see \cite{Liefvendahl2017}.

At each time step, the cell-centred values of the unknowns are interpolated to obtain the values at the face centres.
The scheme used to perform the interpolation has a profound effect on the numerical dissipation of the overall algorithm.
A common scheme to use in conjunction with traditional, wall-resolved LES (WRLES) is linear interpolation using the values in the centres of the cells sharing the face, see e.g.~\cite{Temmerman2003, Frohlich2005, Breuer2009, Bentaleb2012}.
This scheme is second-order accurate but not bounded when applied to interpolation of the convective fluxes.
Using it for this purpose, therefore, leads to the introduction of numerical oscillations.
This does not possess a problem in the case of WRLES due to the associated small grid-cell size, but for WMLES these oscillations can potentially contaminate the solution significantly.
Moreover, based on the experience of the authors, when an unstructured grid is used, divergence of the entire simulation can be expected.
Therefore, besides for linear interpolation, the linear-upwind stabilized transport (LUST) scheme~\cite{Weller2012, Martinez2015} is also considered as a candidate for discretising the convective term in~\eqref{eq:les}.
This scheme computes the face-centred value using a weighted average of the value obtained by linear interpolation (75\%), and that obtained using a second-order upwind scheme (25\%).
The accuracy of the scheme is thus second-order, but the oscillations coming from linear interpolation are smeared out due to the numerical dissipation coming from the upwinding.
For the diffusive cell-face fluxes, linear interpolation can be used without any side-effects.
For numerical integration in time, a second-order implicit backward-differencing method as described in~\cite{Jasak1996} is used.  
The PISO algorithm~\cite{Issa1986} is used for pressure-velocity coupling, with three pressure correction iterations performed at each time step.

An important factor in LES of wall-bounded turbulent flows is employing proper boundary conditions for the governing equations at the wall. 
As discussed in \cite{Bose2014, Sagaut2005}, only upon sufficient reduction of the filter width adjacent to the wall, can the boundary conditions of the unfiltered quantities  be used for the corresponding filtered ones. 
In the context of implicitly-filtered LES, this is equivalent to having sufficiently fine meshes adjacent to the wall, which  is the case for wall-resolving LES. 
Otherwise, it is required to implement a special type of wall treatment, i.e.~a wall model, in order to compute and impose the correct boundary conditions. 
As noted in the introduction, here the focus is on a treatment known as wall-stress modelling, which is discussed in the next section.

\section{Modelling the wall shear stress}
\label{sec:wss}

\subsection{Overview of the approach}
\label{sec:wss_overview}
The wall modelling, and how it is connected to the wall stress in the finite volume framework is now described.
The algebraic and ODE-based models of Sections~\ref{sec:wss_algebraic} and~\ref{sec:wss_ode} can, however, also be integrated into other numerical frameworks.

Consider a finite volume cell with a face of size $S_w$ adjacent to the wall.
For simplicity, assume that the wall lies in the $x_1$-$x_3$ plane of a Cartesian coordinate system, and $x_2$ points in the wall-normal direction into the fluid domain.
Then, for the wall-parallel components, the integral form of the momentum equation (\ref{eq:les}) reads,

\begin{equation}\label{eq:lesfvm}
\diff{}{t}\int_V \olu_i \,\mbox{d}V + \oint_S \olu_i \olu_j n_j \mbox{d}S =
- \frac{1}{\rho} \oint_S \olp n_i\,\mbox{d}S + \oint_S \tau_{ij}n_j\,\mbox{d}S.
\hspace{1cm} (i=1,3)
\end{equation}

Consider the last term in the above equation, which accounts for the effects of the subgrid scale and viscous momentum fluxes.
In the finite volume framework, the surface integral is decomposed into a sum over the faces of the considered cell.
As a result, the contribution to the sum from the face at the wall is obtained as,

\begin{equation} \label{eq:tau_w}
\int_{S_w} \tau_{ij}n_j \mbox{d}S= -\int_{S_w} \tau_{i2}\,\mbox{d}S \approx \overline \tau_{w,i2}S_w.\hspace{1cm} (i=1,3)
\end{equation}
The quantities $\overline \tau_{w, 12}$ and $\overline \tau_{w, 32}$ are the two wall-parallel components of the filtered wall shear stress vector.
The magnitude of this vector is referred to as $\oltau_w$.

\begin{figure}[htp!]
    \centering
    \includegraphics[trim={0 0.65cm 0 0},clip]{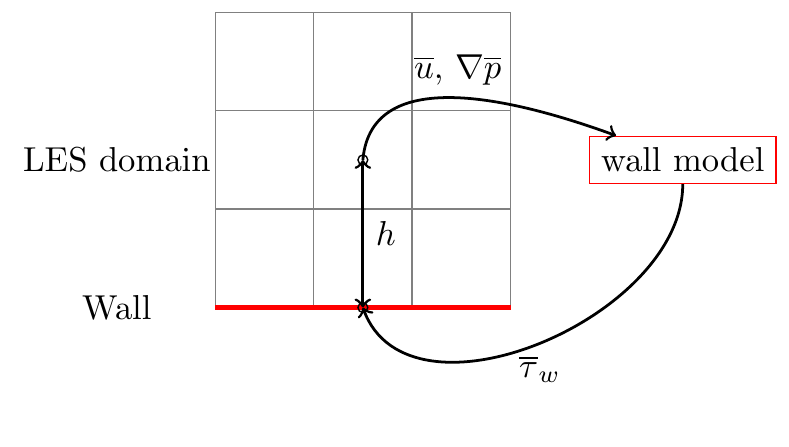}
    \caption{Mode of operation of a wall-stress model. Quantities from the simulated flow are sampled from a cell at a distance $h$ from the wall and serve as input to the wall model. The wall model estimates $\oltau_w$, which is then enforced at the centre of the corresponding wall face.}
    \label{fig:wm}
\end{figure} 

The idea of wall-stress modelling is to introduce a procedure for estimating and enforcing the correct local value of $\oltau_w$ at each wall face.
Schematically, see also Figure~\ref{fig:wm}, this can be summarised as applying the following three steps at each boundary face, at each time step of the simulation.
\begin{description}
    \item[Step 1] The values of $\olu$, $\olp$, or quantities derived from them (e.g.~the pressure gradient) are sampled from a cell centre in the LES domain, located at some wall-normal distance from the wall, $h$.
    The sampled values serve as input to the wall model.
    
    \item[Step 2] From the sampled input values, the local value of  the filtered wall shear stress $\overline \tau_w$ is computed using the wall model.
    
    \item[Step 3] The computed $\overline \tau_w$ is  enforced at the given face centre.    
\end{description}

The first two steps of the algorithm are given attention in the sections below.
Here, the discussion continues with considering Step 3.
Employing the no-slip condition for $\olu_i$, the standard finite volume approximation of $\tau_{i2}$ at the wall gives the following relation,
\begin{equation}
  \overline \tau_{w,i2} = (\nu + \nu_{\text{sgs}})_f \frac{\overline u_{i,P}}{\Delta x_2}. \quad(i=1,3)
\end{equation}
The sub-script $P$ implies evaluation in the centre of the wall-adjacent cell, the sub-script $f$ evaluation in the centre of the wall face, and $\Delta x_2$ is the wall-normal distance between these two points.
The correct value of $\overline \tau_w$ is then enforced if the following value of $\nu_{\text{sgs}}$ is set at the wall face,
\begin{equation}\label{eq:nu_corr}
  \nu_{\text{sgs}} = \frac{\overline \tau_w}{\left[
      \left( \overline u_{1,P}/\Delta x_2\right)^2 +
      \left( \overline u_{3,P}/\Delta x_2\right)^2
      \right]^{1/2}} - \nu.
\end{equation}
Note that here the no-slip condition for $\olu_i$ is essentially a part of the wall modelling procedure, and not a physical boundary condition, as it is in the case of a wall-resolved simulation.
Also, an implicit assumption is that the local wall shear stress is aligned with the wall-parallel velocity in the centre of the wall-adjacent cell.

\subsection{Algebraic models}
\label{sec:wss_algebraic}

The study of (equilibrium) TBLs  has led to several functional relationships  of the type $\mean{u}^+(x_2^+)$ being proposed --- so-called ``laws of the wall''.
Here, $\mean{u}^+ = \mean{u}/\mean{u_\tau}$, $x_2^+=x_2\mean{u_\tau}/\nu$ is the normalized distance to the wall, and the brackets $\mean{\cdot}$ indicate averaging (in time and/or ensemble).

The most well-known law of the wall is the log-law, which is valid approximately between $x_2^+ \approx 30$ and $x_2/\delta \approx 0.3$,~\cite{Pope2000}.
Other laws, matching the laminar sublayer and the log-law, have also been proposed.
In particular, in this work, the following two will be used.
Spalding's law~\cite{Spalding1961},
\begin{equation}
    \label{eq:spalding}
	x_2^+ = \mean{u}^+ +e^{-\kappa B}\left[e^{\kappa \mean{u}^+} -1 -\kappa \mean{u}^+ - \frac{1}{2}(\kappa \mean{u}^+)^2 -\frac{1}{6}(\kappa \mean{u}^+)^3  \right],
\end{equation}
where the model parameters have values $\kappa = 0.4$ and $B = 5.5$ .
And Reichardt's law~\cite{Reichardt1951},
\begin{equation}\label{eq:Reichardt}
\mean{u}^+= \frac{1}{\kappa} \ln \left(1+\kappa x_2^+ \right) 
+ C \left(1-e^{(-{x_2^+}/{B_1})}-\frac{x_2^+}{B_1}e^{(-{x_2^+}/{B_2})}  \right) \,,
\end{equation} 
with model parameter values $\kappa=0.4$, $C=7.8$, $B_1=11$, and $B_2=3$. 

Algebraic wall models are based on the assumption that a law of the wall is also valid for~$\olu$, the instantaneous filtered velocity sampled from the LES solution.
Then, given $\olu$, $x_2$ and $\nu$,  the law provides a non-linear algebraic equation, which can be solved to obtain  the local filtered wall shear stress,~$\overline{\tau}_w$.

While there is no reason to expect that a law developed for the mean velocity will perfectly capture the behaviour of $\olu$, it can be argued that it can provide a good approximation.
The main line of reasoning found in the literature, see e.g.~\cite{Piomelli2002, Larsson2016}, is that at large Re-numbers the resolution of a grid constructed to resolve only the outer layer  will be so coarse compared to the characteristic length scale of the inner layer $\delta_\nu$ that the inner layer dynamics can be considered in the mean sense.
As a consequence, applying relationships valid for averaged quantities, such as~\eqref{eq:spalding} and~\eqref{eq:Reichardt}, becomes justifiable.

\subsection{Integrated formulation of algebraic models}
\label{sec:wss_algebraic_int}
In~\cite{Werner1991}, Werner and Wengle introduced an algebraic wall model based on the following power-law for the mean velocity profile,

\begin{equation}
\label{eq:wernerwengle}
\mean {u}^+ = \begin{cases} x_2^+ &\mbox{if } x_2^+ \leq 11.81, \\
A(x_2^+)^B & \mbox{otherwise,} \end{cases} 
\end{equation}
where $A=8.3$ and $B=1/7$.
Instead of directly applying this law of the wall to compute $\oltau_w$ in the manner described in the section above, they used it to derive an expression connecting $\oltau_w$ to the wall-normal average of $\mean{u}^+$ across the interval $[0, h]$.
In the case of \eqref{eq:wernerwengle}, this expression turns out to be explicit, thus avoiding the need to solve a non-linear equation.
More generally, assume that $\olu$ is sampled from a cell extending across the interval $[h_1, h_2]$ in the wall-normal direction, with the corresponding cell centre located at \mbox{$h=(h_1 + h_2)/2$}.
Let $\mean{u}^+ = L(\mean{u_\tau}, x^+_2, q)$ be some law of the wall, e.g.~\eqref{eq:Reichardt} or \eqref{eq:wernerwengle}, where $q$ are the model parameters.
Averaging over the wall-normal interval $[h_1, h_2]$ leads to the following relationship,
\begin{equation}
\label{eq:integrated_law}
\frac{1}{h_2-h_1}\int_{h_1}^{h_2} \langle u \rangle \text{d} x_2 = \frac{\langle u_\tau \rangle}{h_2-h_1}\int_{h_1}^{h_2} L(\langle u_\tau \rangle, x^+_2, q) \text{d} x_2.
\end{equation}

Recall that in the collocated finite volume method, the stored cell-centred values of the unknowns approximate their averaged values across the volumes of the corresponding cells.
Therefore, it can be argued that $\olu$ more accurately approximates the left-hand-side of~\eqref{eq:integrated_law}, rather than  the point-wise value of $\mean{u}$ at $h$.
Employing this assumption and exchanging $\mean{u_\tau}$ to the unknown local filtered friction velocity, $\bar u_\tau$, leads to
\begin{equation}
\label{eq:integrated}
(h_2-h_1)\bar u - \bar u_\tau \int_{h_1}^{h_2}L( \olu_\tau, x^+_2, q) \text{d} x_2 = 0.
\end{equation}
If the intgral in the above equation can be taken analytically (which is the case for the laws of Riechardt,~\eqref{eq:Reichardt} and Werner and Wengele,~\eqref{eq:wernerwengle})
a non-linear algebraic equation for $\bar u_\tau$ is obtained.
Thus, an algebraic wall model based on a law of the wall explicitly relating $\mean{u}^+$ to $x_2^+$, can be formulated in two ways.
The first formulation, here referred to as \textit{standard}, is based on directly using the law of the wall as a non-linear equation to compute $\overline{\tau}_w$, as discussed in Section~\ref{sec:wss_algebraic}. 
The second formulation, referred to as \textit{integrated}, is obtained by first taking a wall-normal average, as presented above.

In~\cite{Temmerman2003}, the authors report results from WMLES of the flow over periodic hills using a finite volume-based code.
The performance of the standard and integrated formulations of the same algebraic wall models was compared.
In particular, two laws of the wall were considered: the power law~\eqref{eq:wernerwengle} and a ``two-layer log-law'' (see~\cite{Temmerman2003} for details).
In all simulations, the velocity was sampled from the wall-adjacent cell.
For both considered laws, the integrated formulation yielded more accurate results.

\subsection{ODE-based models}
\label{sec:wss_ode}

The common starting point for deriving wall-stress models based on differential equations are the turbulent boundary layer equations (TBLE),~\cite{Cabot1999, Wang2002, Piomelli2002, Piomelli2008},

\begin{align}
    \label{eq:tble}
    &   \pdiff{}{x_2} \left[ (\nu + \nu_t)\pdiff{{\mean{u_i}}}{x_2}\right] = F_i,
\end{align}    
where
\begin{align}
    \label{eq:fullF}
    & F_i = \frac{1}{\rho}\pdiff{\mean{p}}{x_i} + \pdiff{\mean{u_i}}{t} + \pdiff{}{x_j}(\mean{u_i}\mean{u_j}).
\end{align}
Here, $i=1,3$, and the wall-normal velocity can be found via the continuity equation.
In the formulation above, the Boussinesq assumption concerning the form of the Reynolds stress tensor has been made.

The TBLE can be solved to obtain the values of $\oltau_w$ and, hence, constitute a PDE-type wall model.
In general, the PDEs should be solved on a separate, ``embedded'', three-dimensional grid spanning the region between the wall and $h$.
For complex geometry, this task is far from trivial to do automatically, and delegating it to the user means a significantly more laborious process of setting up the simulation.
As it was mentioned in the introduction, there is also currently no consensus regarding whether PDE-based models are necessarily more accurate than simpler ones~\cite{Larsson2016}.
Consequently, PDE-models are not further considered here.

An alternative approach is to instead sample one or several of the terms in~\eqref{eq:fullF} from a single point in the LES solution.
In that case, $F_i$ becomes a constant source term, and~\eqref{eq:tble} takes the form of an ODE, which can be solved for each wall-adjacent cell face.
In fact, it becomes possible to integrate~\eqref{eq:tble} analytically, see~\cite{Wang2002}, leading to
\begin{align}
    \label{eq:wmint}
    &   \mean {\tau_{w,i}} =\left(\mean{u_i}\vert_h -F_i \int^h_0 \frac{x_2 }{\nu + \nu_t}\text{d}x_2  \right) \bigg / \int^h_0 \frac{\text{d}x_2}{\nu + \nu_t} , \quad (i=1,3),
\end{align}
and to the following expression for the magnitude of the mean wall shear stress
\begin{align}
    \label{eq:odetauw}
    &   \mean{\tau_w} =  \left( \mean{u_i}\vert_h \mean{u_i}\vert_h + F_iF_i\left(\int^h_0 \frac{x_2 }{\nu + \nu_t}\text{d}x_2\right)^2  - 2\mean{u_i}\vert_h F_i \int^h_0 \frac{x_2 }{\nu + \nu_t}\text{d}x_2  \right)^{1/2}\bigg / \left \vert \int^h_0 \frac{\text{d}x_2}{\nu + \nu_t}\right \vert .
\end{align}
To find $\mean{\tau_w}$, it is thus sufficient to numerically compute the integrals in the above expression.
Similarly to how algebraic models are used, it is possible to employ~\eqref{eq:odetauw} to compute $\oltau_w$ given $F_i$ and $\olu_i$ sampled from the simulation.

What remains to be discussed is the choice of model for $\nu_t$.
Since variation in only one spatial dimension is allowed, the common approach is to employ a mixing length type of turbulence model.
A popular choice,~\cite{Cabot1999, Wang2002, Kawai2013}, is an expression based on the mixing length model coupled with the van Driest damping function near the wall~\cite{vanDriest1956},
\begin{equation}
    \label{eq:johnsonking}
    \nu_t = \nu \kappa x^+_2 \left( 1 - \exp(-x_2^+/A)\right)^2,
\end{equation}
where the values used for the parameters are $\kappa = 0.4$ and $A = 17.8$.
Note that~\eqref{eq:johnsonking} depends on the value of~$u_\tau$, which is not known.
Fixed point iteration using~\eqref{eq:odetauw} and~\eqref{eq:johnsonking} until convergence is therefore used.

In~\cite{Balaras1996}, a similar $\nu_t$ model is proposed that is further developed by Duprat et al~\cite{Duprat2011} to better work in simulations of flows with separation.
To that end, the authors employ the velocity scale introduced in~\cite{Manhart2008}, \mbox{$u_{\tau p} = (u_\tau^2 + u_p^2)^{1/2}$}, where $u_p = |\nu/\rho (\partial p/\partial x)|^{1/3}$ and $x$ is the streamwise coordinate.
The non-dimensional parameter $\alpha = u^2_\tau/u^2_{\tau p}$ is defined to measure the relative strength of the shear stress and the streamwise pressure gradient.
Note that unlike $u_\tau$, $u_{\tau p}$ does not become zero at separation and reattachment points.
Defining $x_2^* = x_2u_{\tau p}/\nu$, the expression for the turbulent viscosity is

\begin{equation}
\label{eq:duprat}
\nu_t = \nu \kappa x_2^*\left[\alpha + x_2^*(1-\alpha)^{3/2} \right]^\beta \left[1 - \exp \left(-\frac{x_2^*}{1 + A\alpha^3} \right) \right]^2,
\end{equation}
where $A=17$ and $\beta = 0.78$ are model parameters.

Within this ODE-model framework, the key question that needs to be addressed is the treatment of the components of the source term, $F_i$.
Ideally, it is supposed to model the combined effect of convective, transient, and pressure gradient terms.
Commonly, however, only some of the terms are explicitly taken into account.
The simplest choice, $F_i=0$, leads to ODE-models that are essentially equivalent to using a law of the wall.
Choosing,
\begin{equation}
\label{eq:Fpressure}
F_i= \frac{1}{\rho}\pdiff{\mean{p}}{x_i},
\end{equation}
has received attention, due to the role of the adverse pressure gradient in the process of flow separation.
In~\cite{Duprat2011}, this treatment of $F_i$ coupled with the $\nu_t$ model defined by~\eqref{eq:duprat}, was shown to give improved results for periodic flow over a hill, as compared to an algebraic model based on Spalding's law.
In~\cite{Wang2002}, however, the results obtained by only considering the pressure gradient term were unsatisfactory.
In~\cite{Larsson2016}, the authors argue that one must either include all the terms in~\eqref{eq:fullF} or neither of them.
In summary, the correct treatment of $F_i$ in ODE-type models remains an open question.

\subsection{Grid resolution requirements}
\label{sec:wss_numerical}

By definition, WMLES aims at accurately resolving the turbulence in the outer layer, where the characteristic length scale is $\delta$.
The question is what grid resolution (with respect to $\delta$) is sufficient to achieve that requirement.
Let $n_0$, be the amount of cells per $\delta^3$-cube.
Appropriate values of $n_0$ can be found in the literature.
In~\cite{Chapman1979}, $n_0 = 2\,500$ is recommended.
In~\cite{Spalart1997}, numbers in the range from $8\,000$ to $27\,000$ are considered sufficient.
More recently in~\cite{Larsson2016}, the value of $\approx 6\,500$ is recommended.
Further testing of different values of $n_0$, across simulations of different flows, appears to be necessary to make the recommendations more precise.

The authors of~\cite{Kawai2012} criticize the common practice of using the grid point closest to the wall for sampling wall model input from the LES.
In short, their argument is as follows.
The size of the energetic motions in the logarithmic layer is proportional to the wall-normal distance, $L \sim Cx_2$.
To resolve $L$, some number of grid points $N$ is needed, so the grid size should be $\Delta x_2 \lesssim L/N = Cx_2/N$.
At the first off-the-wall grid point, the distance from the wall is $h$, and so is the resolution of the grid.
As a result, the turbulent motion is properly resolved only if $ C \gtrsim N$, which the authors of~\cite{Kawai2012} deem to be highly unlikely due to damping of the turbulent eddies by the wall.
These arguments are supported by flat-plate TBL simulations, with larger $h$ giving better predictions of the mean wall shear stress.
In particular, no LLM is observed contrary to when the point next to the wall is used.
Similar results were obtained in~\cite{Lee2013}, and more recently in~\cite{Frere2017}.

\section{Description of the software}
\label{sec:library}

\subsection{Current WMLES capabilities of OpenFOAM}
This section details the functionality of a new library for WMLES, implemented using \ttt{OpenFOAM} technology.
But to put the novel features of the code into context, the wall-stress modelling capabilities of the standard \ttt{OpenFOAM} library are briefly reviewed first.
For a more detailed discussion, see~\cite{Liefvendahl2017,DeVilliers2006}.

Two algebraic wall-stress models are included.
One is based on Spalding's law of the wall~\eqref{eq:spalding} and the other on the log-law. In the latter, a modified length scale is introduced in order to account for wall-roughness.
The user is given limited control over the model parameters.
In particular, the distance to the sampling point, $h$, is fixed to be the distance to the centre of the wall-adjacent cell.
As discussed in Section~\ref{sec:wss_numerical}, this may lead to sub-optimal results.

The enforcement of $\oltau_w$ is performed according to~\eqref{eq:nu_corr}.
This is implemented by having the wall models' classes inherit from the base class for Dirichlet-type boundary conditions.
This effectively makes any wall model just another type of boundary condition among many others available in \ttt{OpenFOAM}.
This is an excellent design choice because it decouples the wall model code from the code of any particular LES solver.
So, in principle, any future \ttt{OpenFOAM} solver for the LES equations~\eqref{eq:les} (e.g. based on a different pressure-velocity coupling algorithm) can make use of the wall model.
For this reason, the same design is adopted for the new WMLES library.

For completeness, it should be noted that \texttt{OpenFOAM} also has extensive support for DES and related methods. 
As it was mentioned in the introduction, they can also be used for conducting WMLES, but the methodology is entirely different from wall-stress modelling, which is the focus of this work.

\subsection{Design and programming interface}
\label{sec:design}

One of the main goals of the new library is to introduce a convenient class structure for implementing wall models, which would allow avoiding unnecessary code duplication.
To that end, an effort was made to decouple the implementation of the wall models themselves and that of  related functionality and concepts, such as root finding algorithms, laws of the wall, eddy viscosity models, field sampling, etc.
Further, the programming interface is designed in a way that allows implementing new wall models by inheriting from the base class of the appropriate model type and re-implementing one or two key virtual functions.

\begin{figure}
    \centering
    {\includegraphics[trim=0 0 0 0,clip ]{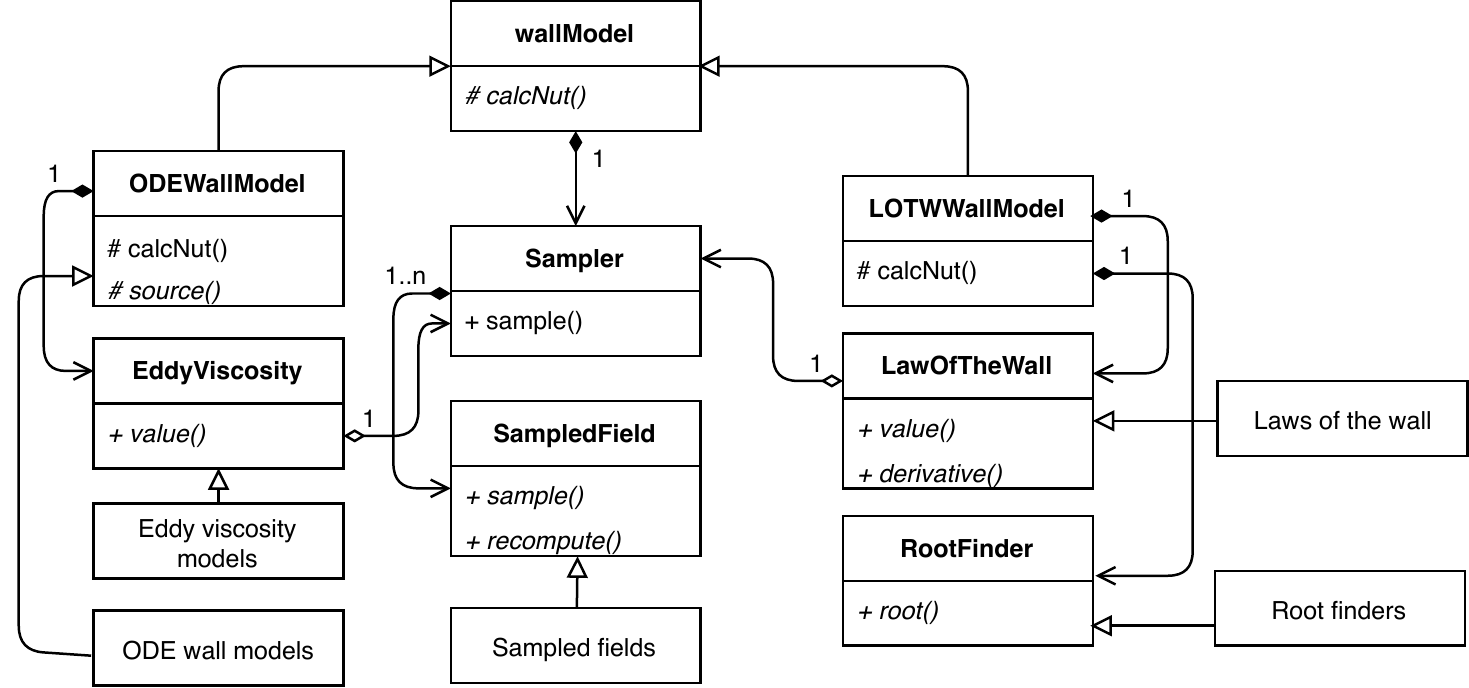}}
    \caption{Universal modelling language (UML) diagram of the class structure of the library.}
    \label{fig:classes}
\end{figure}

A diagram of the adopted class structure is shown in Figure~\ref{fig:classes}.
The base abstract class \ttt{wallModel} inherits from the base class for Dirichlet boundary conditions (the latter not shown in the figure) and implements common functionality for all wall models.
An abstract method, \texttt{calcNut}, is defined to compute the updated values for $\nu_\text{sgs}$ at the wall.
All the wall models are thus defined by their implementation of this method.

Two classes inherit from \ttt{wallModel}: \ttt{ODEWallModel} and \ttt{LOTWWallModel}, corresponding to ODE-based and algebraic (law of the wall-based) models.
All ODE models use equations~\eqref{eq:odetauw} and~\eqref{eq:nu_corr} to compute $\nu_\text{sgs}$, and differ only in the treatment of the source term approximating the right-hand-side of the TBLE equations~\eqref{eq:tble}, i.e.~$F_i$.
Therefore, the \ttt{ODEWallModel} class implements~\eqref{eq:odetauw} and~\eqref{eq:nu_corr} in \ttt{calcNut}, whereas computing $F_i$ is performed in a separate method, \ttt{source}, which is declared as abstract.
Each individual ODE-based wall model inherits from~\ttt{ODEWallModel} and implements \ttt{source}, thus defining its specific treatment of the source term.
A separate class hierarchy is set up for different $\nu_t$ models, the base class being \ttt{EddyViscosity}.
The inheriting classes implement the \ttt{value} method, which returns the values of $\nu_t$ at the location of the nodes of the embedded 1D grid.
This allows to avoid introducing a new ODE-based wall model for each new $\nu_t$ model, thus providing an example of how the structure of the code leads to a reduction in code duplication.

A similar approach is used for the algebraic wall model, associated laws of the wall and equation root finders.
The \ttt{LOTWWallModel} class holds a reference to an object of the \ttt{LawOfTheWall} class, which defines two abstract methods implemented by all inheriting law of the wall classes: \ttt{value} and \ttt{derivative}.
The former implements the relationship between $h$, $\nu$, the necessary quantities sampled from the LES solution, and the value of $\olu_\tau$, as defined by the law.
The latter does the same thing, but for the derivative of this relationship with respect to $\bar u_\tau$.
Note that both standard and integrated algebraic wall-models can be incorporated into this structure.
The \ttt{value} and \ttt{derivative} functions are sent by the \ttt{LOTWWallModel} class to a \ttt{RootFinder} object, which uses them to iteratively find $\bar u_\tau$.

To handle the sampling of the input data to the model from the LES domain, the \ttt{Sampler} class is introduced.
At the beginning of the simulation, the \ttt{Sampler} reads the values of $h$ from disk and finds the indices of the corresponding cells that will be used for sampling.
Additionally, it computes the values of $h_1$ and $h_2$ for each cell, as required for integrated algebraic wall models, see~\eqref{eq:integrated}.
The \ttt{Sampler} also holds a list of pointers to objects derived from the \ttt{SampledField} class, which correspond to different types of input to the wall model.
At each time step the \ttt{sample} method of the \ttt{Sampler} is called, which triggers the sampling of all the \ttt{SampledField} objects in the list.
Note that leaving the access to the \ttt{Sampler} only to the \ttt{wallModel} class would make it difficult to determine what fields have to be sampled at run time.
For instance, an equilibrium ODE model does not need to use the pressure gradient, however, if coupled with the $\nu_t$ model defined by equation~\eqref{eq:duprat}, the pressure gradient is needed.
Therefore, full access to the \ttt{Sampler} is given to the \ttt{EddyViscosity} and \ttt{LawOfTheWall} classes, allowing them to add \ttt{SampledField} objects to its list of sampled fields, using the \ttt{addField} method.

Currently, three classes inheriting from \ttt{SampledField} are present in the library, corresponding to the velocity, pressure gradient and the wall-normal gradient of the velocity.
All three quantities are projected onto the wall-parallel direction.
If a new model that needs some other field to be sampled is added, the corresponding class should be implemented.
The \ttt{recompute} method is used to define how the field should be calculated given the LES solution, with the possibility to employ any of the differential and algebraic operators defined in \ttt{OpenFOAM}.
The \ttt{sample} method is used to define which values of the computed field are sampled and further manipulated (e.g.~projected).
It should be noted that all sampled fields are added to \ttt{OpenFOAM}'s object registry, and are written to disk along with the other fields.

\subsection{User interface, included wall models and features}
\label{sec:included_wm}

From the user's side, the set-up of the wall model occurs in the \ttt{nut} initial data file, which holds the values of~$\nu_\text{sgs}$ and also defines its boundary conditions.
The choice of wall model and its configuration is performed by setting up the appropriate boundary condition for \ttt{nut} at wall boundaries.
Based on parsing these dictionaries, the appropriate classes are constructed at run time by \ttt{OpenFOAM} using a run-time selection mechanism.

A typical configuration dictionary is shown in Listing~\ref{lst:wm}.
The wall model is chosen using the \ttt{type} keyword.
The parameters of sub-components of the wall model are defined in associated sub-dictionaries.
As demonstrated in Listing~\ref{lst:wm}, in the case of \ttt{LOTWWallModel} the user can choose what root finding method and law of the wall to employ.
Most parameters have associated default values, which the model falls back to when user input is not provided.

\begin{Listing}
\RawFloats
\begin{lstlisting}[caption={A configuration dictionary for the \ttt{LOTWWallModel} for some boundary patch called \ttt{bottomWall}. Selection of the algebraic model based on Spalding's law. The value of $\kappa$ defined by the user, the default value 5.5 will be used for $B$.}]
bottomWall
{    
    type        LOTWWallModel;
    value       uniform 0;
    RootFinder
    {
        type    Newton;
        maxIter 10;
    }
    Law
    {
        type    Spalding;
        kappa   0.4;
    }
}
\end{lstlisting}
\label{lst:wm}
\end{Listing}
The user is also expected to provide the values of $h$ for each boundary where wall modelling is applied.
This is done in a separate scalar field, \ttt{h}, which all the wall models read.
A single $h$ value can be provided for the whole boundary, or a list of values, corresponding to each boundary face.
The latter can be useful when the thickness of the TBL, $\delta$, changes significantly over a boundary patch and a constant value of $h/\delta$ across the boundary is desired.
For convenience, the value $h = 0$ is reserved to correspond to sampling from the centre of the wall-adjacent cell.
Note that, in a parallel setting, the wall face and the appropriate sampling cell may end up residing on different processor-domains.
Such cases are detected by the algorithm and handled by falling back to sampling from the wall-adjacent cell.
For convenience, the wall model writes out a field, \ttt{samplingCells}, where the cells selected for sampling are marked using the index of the associated wall-patch.

All the models and methods included in the library are summarised in Table~\ref{tab:features}.
The corresponding values for the \ttt{type} keyword that should be used in the configuration dictionary are also provided.
Five algebraic wall models are currently present.
The first is based on Spalding's law~\eqref{eq:spalding}.
Two are based on the power law introduced by Werner and Wengle~\eqref{eq:wernerwengle}, the standard and integrated formulation, respectively.
Similarly, two models are based on Reichardt's law~\eqref{eq:Reichardt}.
The values of all model parameters (e.g. $\kappa$ and $B$ for Spalding's law) can be configured by the user.

Two root finders are implemented for solving the associated algebraic equations.
One based on the Newton-Raphson method and the other on the bisection method.
The latter is meant for use with laws of the wall for which a derivative cannot be meaningfully defined.
It is possible for the user to configure the tolerance and maximum number of iterations that the root finders are allowed to take.

Two ODE-based models are included in the library.
The first one, \ttt{EquilibriumODEWallModel}, assumes~$F_i$ in~\eqref{eq:fullF} to be identically zero.
The other one, \ttt{PGradODEWallModel}, assigns the source term the value of the gradient of~$\olp$ projected onto the wall-parallel direction.
For numerically computing the integrals present in~\eqref{eq:odetauw} the trapezoidal rule is used.
The number of points used for the integration is defined by the user.
Two models for computing $\nu_t$ are available, defined in equations~\eqref{eq:johnsonking} and~\eqref{eq:duprat} respectively.

In the course of the simulation, the obtained values of $\oltau_w$ are stored in the \ttt{wallShearStress} field, which is saved to disk along with the other unknowns.
The \ttt{fieldAverage} function object built into \ttt{OpenFOAM} can be used to obtain the mean $\oltau_w$ and its standard deviation.

\begin{table}[htp]
\begin{tabular}{lll}
                            & \textbf{Model or method}                            & \textbf{Name in library}       \\ 
\textbf{Wall models}        & Algebraic                                           & \ttt{LOTWWallModel}            \\
                            & ODE-based, $F_i = 0$                   & \ttt{EquilibriumODEWallModel} \\
                            & ODE-based, $F_i = \frac{1}{\rho}\frac{\partial \olp}{\partial x_i}$ & \ttt{PGradODEWallModel}        \\ \hline
\textbf{Laws of the wall}   & Spalding's                                          & \ttt{Spalding}                 \\
                            & Reichardt's, standard formulation                   & \ttt{Reichardt}                \\
                            & Reichardt's, integrated formulation                 & \ttt{IntegratedReichardt}     \\
                            & Werner and Wengle, standard formulation             & \ttt{WernerWengle}             \\
                            & Werner and Wengle, integrated formulation           & \ttt{IntegratedWernerWengle}   \\ \hline
\textbf{Root finders}       & Newton-Raphson method                               & \ttt{Newton}                   \\ 
                            & Bisection method                                    & \ttt{Bisection}                \\ \hline
\textbf{Models for $\nu_t$} & Mixing length with van Driest damping               & \ttt{VanDriest}              \\ 
                            & Model of Duprat et al,~\cite{Duprat2011}             & \ttt{Duprat}                  
\end{tabular}
\caption{The models and methods included in the library and the corresponding values that should be used for the \ttt{type} keyword in the configuration dictionary.}
\label{tab:features}
\end{table}
\section{Application to fully-developed turbulent channel flow}
\label{sec:channel}
The developed library was used for WMLES of fully-developed turbulent channel flow.
The performed simulations evaluate the predictive accuracy of several algebraic wall models as well as how the results are influenced by other modelling choices.

Both here and in the next section concerning the backward-facing step, a Cartesian coordinate system~$(x, y, z)$ is used, with the three axes corresponding to the streamwise, wall-normal and spanwise directions, respectively.
The corresponding components of the filtered velocity are denoted  $u$, $v$, and $w$.
Further, for all reported quantities the overbar is dropped to simplify the notation. 

\subsection{Case set-up}
Fully-developed turbulent channel flow can be simulated using a box-shaped domain, with periodic conditions applied at boundaries that are not walls.
The box lengths in the streamwise, spanwise, and wall-normal directions are here taken to be $9\delta$, $4\delta$, and $2\delta$, respectively, with $\delta = 1$ m denoting the channel half-height.
To ensure that the domain is sufficiently large to accommodate the turbulent structures present in the flow, a selection of the simulations described below was also performed on a domain of a larger size, with no significant difference in the results observed.

Channel flow can be fully defined by the value of $\rey_b = U_b \delta/\nu$, where \mbox{$U_b = 1/\delta \int_0^\delta \mean u \text{d}y$} is the bulk velocity.
Here, $\rey_b = 125\, 000$ is considered.
To fix $U_b$ in the simulations, a source term is introduced into the momentum equation.
The magnitude of the source term is adjusted at every time step in order to enforce the desired value of \mbox{$U_b$}, here $1$ m/s.
Corresponding to the values of $\delta$, $U_b$ and the desired $\rey_b$, the value of $\nu$ is set to \mbox{$8\cdot 10^{-6}$ $\text{m}^2/\text{s}$}.
It is interesting to note that a wall-resolved LES of this flow would require up to $\sim 10^9$ cells using a structured mesh, see~\cite{Rezaeiravesh2017}.

Equivalently, channel flow can be defined by the value of $\rey_\tau  = \mean{u_\tau} \delta/\nu$.
In the employed set-up, $\mean{u_\tau}$  and thus $\rey_\tau$ are outcomes of the simulation.
Based on DNS data~\cite{Lee2015}, the target value of $\rey_\tau$ corresponding to the chosen value of $\rey_b$ is $\rey_\tau \approx 5200$.
Since the correct prediction of $\mean{u_\tau}$ is one of the main objectives of wall modelling, the relative error in this quantity, $\epsilon[\mean{u_\tau}]$, is an important quantitative measure of the accuracy of the performed WMLES.
The error, in $\mean{u_\tau}$ and also other quantities, is here computed with respect to DNS data~\cite{Lee2015}.

The computational domain is meshed with cubic cells.
The resolution of the mesh can be specified as the number of cells used to discretise the channel half-height, $n/\delta$.
The employed value of $n/\delta$ varies from simulation to simulation but is limited to be either 15, 20, 25 or 30.
This corresponds to $n_0$ equal to $3\,375$, $8\,000$, $15\,625$, and $27\,000$, respectively.
As discussed in Section~\ref{sec:wss_numerical} above, all four values are in line with the recommendations found in the literature.

All the simulations use algebraic wall models.
This is motivated by the fact that ODE models based on  $F_i = 0$ are equivalent to algebraic ones in terms of physical assumptions, whereas choosing \mbox{$F_i = 1/\rho \partial  p/\partial x_i$} would have a negligible effect due to the weakness of the pressure gradient driving the flow.
The particular law of the wall employed by the wall model varies and is discussed separately for each set of simulations.

\subsection{Influence of $n/\delta$, $h$, and interpolation scheme for convective fluxes}
\label{sec:channel_parameter_study}
This section presents results from a series of channel flow simulations all of which employ the algebraic wall model based on Spalding's law of the wall~\eqref{eq:spalding} with $\kappa=0.4$ and $B=5.5$ but differ in the choice of other modelling parameters.
In particular, different choices of grid resolution $n/\delta$, distance to the sampling point $h$, and numerical scheme for computing the convective cell-face fluxes are considered.

The simulation campaign consists of 16 simulations, covering all combinations of the following choices of the simulation parameters: $n/\delta \in [15, 20, 25, 30]$, $\text{scheme} \in [\text{linear},\, \text{LUST}]$, $h \in [1^\text{st}, 2^\text{nd}]$.
Here, $h = n^\text{th}$ corresponds to sampling from the center of the $n^\text{th}$ consecutive off-the-wall cell.

\begin{table}[htb!]
\caption{Relative error (in percent) in $\langle u_\tau \rangle$ predictions in channel flow simulations using Spalding's law.}
\centering
\label{tab:channel_utau}
\begin{tabular}{ccccc}
         & \multicolumn{2}{c}{\textbf{Linear}} & \multicolumn{2}{c}{\textbf{LUST}} \\
$\bm{n/\delta}$ & $\bm{h = 1^\text{st}}$                & $\bm{h = 2^\text{nd}}$               & $\bm{h = 1^\text{st}}$               & $\bm{h =2^\text{nd}}$  \\
15       & -4.74                                  & -1.35                         & -15.94                                & 0.74 \\
20       & -6.73                                  & -3.10                         & -16.23                                & 0.00 \\
25       & -7.84                                  & -3.92                         & -16.38                                & -0.62 \\
30       & -8.53                                  & -4.15                         & -16.53                                & -1.25                      
\end{tabular}
\end{table}

The relative error in $\langle u_\tau \rangle$ obtained in the simulations is shown in Table~\ref{tab:channel_utau}.
Perhaps the most interesting result is that accuracy does not improve with the refinement of the grid.
On the contrary, the most accurate prediction of $\langle u_\tau \rangle$ is obtained using the coarsest mesh, with the exception of the case when LUST and $h=2^\text{nd}$ is used, which leads to $\eps [ \mean{u_\tau} ] = 0$ on the $n/\delta = 20$ grid.
Further studies are needed to give an exhaustive explanation of this behaviour.
However, a plausible hypothesis is that on a coarser mesh each sample of the velocity signal better adheres to the employed law of the wall because it corresponds to a spatial average over a larger number of structures on the scale of $\delta_\nu$.
As a result, the wall model performs more accurately.

\begin{figure}[!htb]
\centering
\includegraphics[]{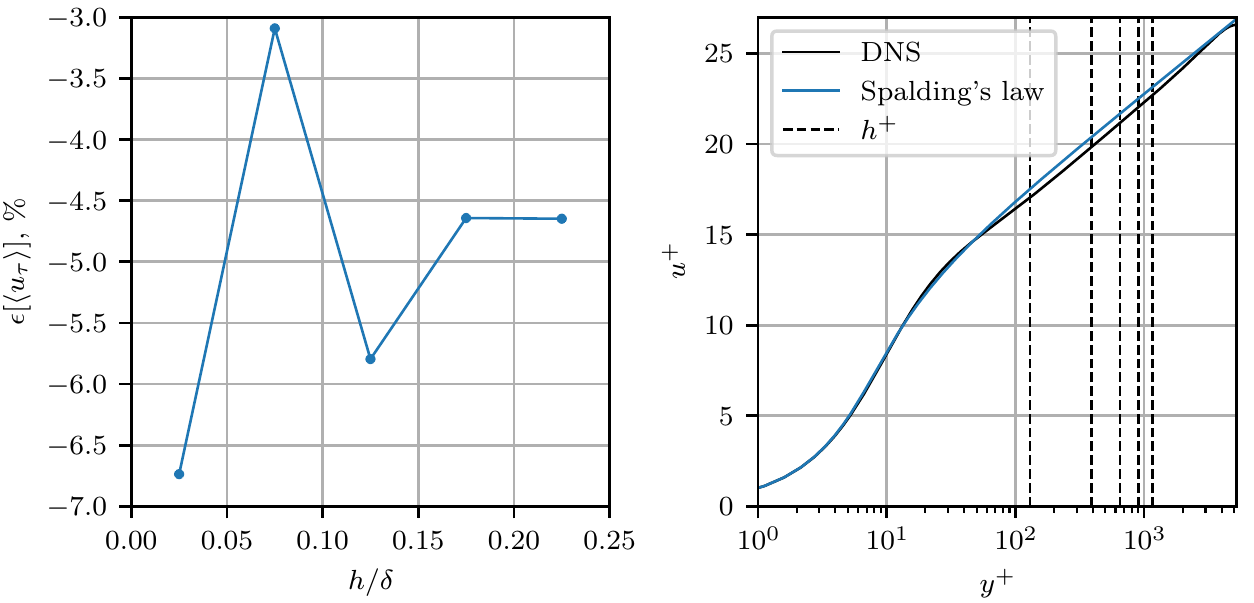}
\caption{\textit{Left}: The obtained relative error in $\langle u_\tau \rangle$ as a function of $h/\delta$. The $n/\delta = 20$ grid and the linear scheme are used in all simulations. \textit{Right:} The location of $h^+$ with respect to the mean velocity profile taken from DNS data~\cite{Lee2015} and that given by Spalding's law~\eqref{eq:spalding}.}
\label{fig:channel_h}
\end{figure}

Another clear outcome is that $h=2^\text{nd}$ leads to an increase of accuracy, as compared to sampling from the wall-adjacent cell.
This corroborates results from previous studies~\cite{Kawai2012, Lee2013, Frere2017}, see the discussion in Section~\ref{sec:wss_numerical}.
It is interesting to see whether further increasing $h$ leads to a further improvement in the accuracy of the results.   
To that end, three additional simulations with $h = 3^\text{rd}$, $4^\text{th}$, and $5^\text{th}$ were performed on the $n/\delta=20$ grid and using the linear scheme.
Note that, as demonstrated in the right plot of Figure~\ref{fig:channel_h}, all five sampling point locations ($h=1^\text{st}$-$5^\text{th}$) are located in the log-law region.
The left plot in Figure~\ref{fig:channel_h} shows the relative error in $\langle u_\tau \rangle$ as a function of $h/\delta$.
It is clear that sampling from the wall-adjacent cell gives the worst accuracy, but increasing $h$ beyond the centre of the second consecutive off-the-wall cell does not result in a further decrease of the error.
A similar trend was observed in~\cite{Kawai2012}.
It should be noted that the magnitude of the error is also affected by the employed law of the wall since its accuracy with respect to the true mean velocity profile varies with $h$, see the right plot in Figure~\ref{fig:channel_h} comparing DNS data and Spalding's law~\eqref{eq:spalding}.

\begin{figure}[!htb]
\centering
\includegraphics[]{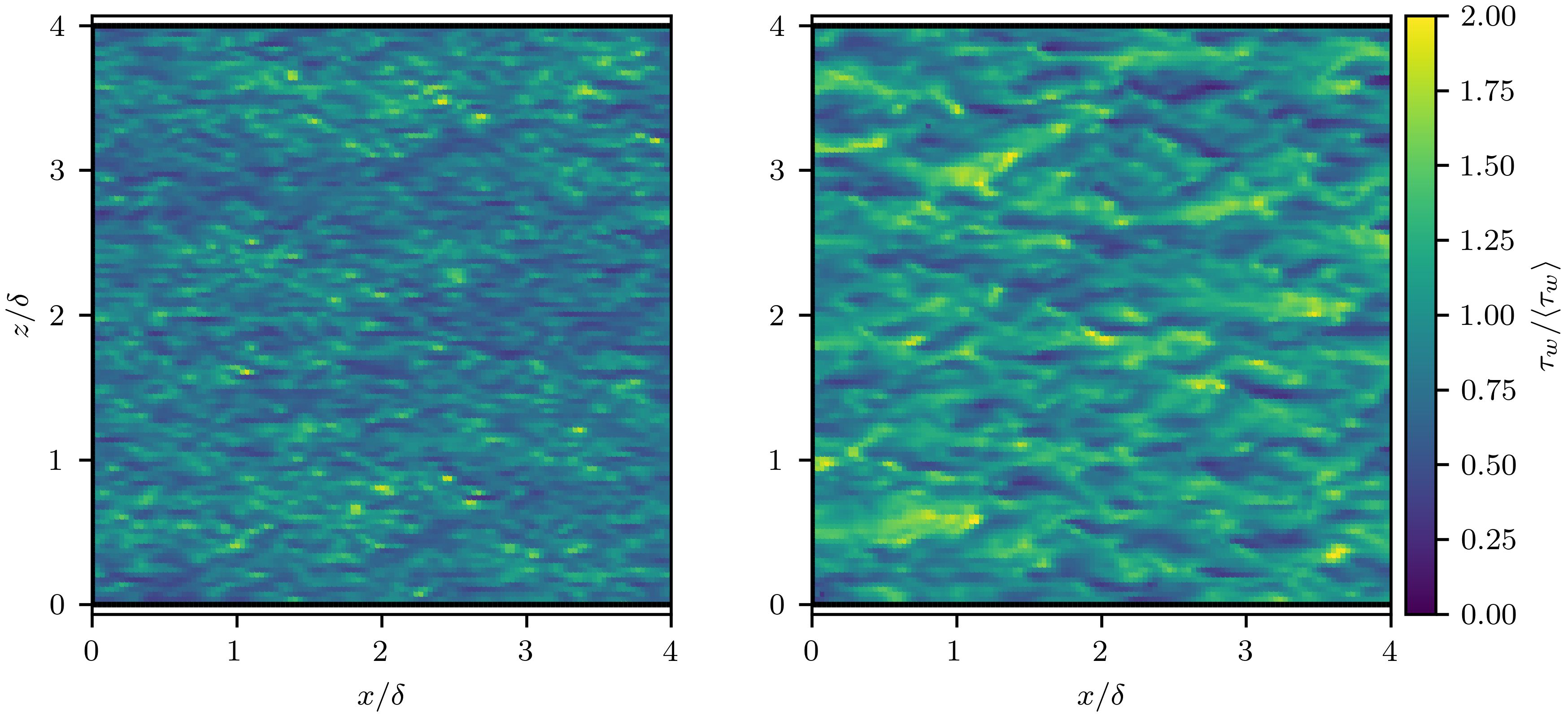}
\caption{The distribution of the normalised instantaneous wall shear stress field over the left half of the bottom wall, taken from a WMLES of channel flow employing the linear scheme (\textit{left}) and the LUST scheme (\textit{right}). Both simulations use the $n/\delta=30$ grid.}
\label{fig:linear_lust}
\end{figure}

Table~\ref{tab:channel_utau} also reveals a large sensitivity of $\eps[\mean{u_\tau}]$ to the choice of the interpolation scheme used for computing the convective cell-face fluxes.
It is noted that while the weight of the diffusive linear upwind scheme in LUST is only 25\%, the amount of numerical diffusion it introduces is significant.
To illustrate this, Figure~\ref{fig:linear_lust} shows instantaneous wall shear stress fields on the bottom wall of the channel, taken from two simulations both using the $n/\delta=30$ grid, but different numerical schemes.
The effect of the extra numerical diffusion in the LUST scheme is evident.
It is, however, not obvious to what extent the more fine-grained variations in $\tau_w$ produced by the linear scheme correspond to resolved turbulent structures and not slight spurious oscillations on a length scale comparable to the grid size.

Analysis of Table~\ref{tab:channel_utau} shows that using LUST results in a stabilising effect on the error in $\langle u_\tau \rangle$ with respect to the choice of $n/\delta$.
For a given choice of $h$, the error difference lies within $1.25$ percentage point across all four considered grid resolutions.
For the linear scheme, the variation reaches $\approx 4$ percentage points.
By contrast, the accuracy of the linear scheme seems to be more stable with respect to the choice of $h$.
Using $h = 2^\text{nd}$ reduces the error by $\approx 1$-3 percentage points, whereas for the LUST scheme $\eps[\mean{u_\tau}]$ is as high as $16.54$\% when $h = 1^\text{st}$ is used but drops to $\approx 1$\% when $h = 2^\text{nd}$ is employed.
Overall, the best results in terms of $\mean{\tau_w}$ are obtained using the LUST scheme and sampling from the second off-the-wall cell centre.

\begin{figure}[!htb]
\centering
\includegraphics[]{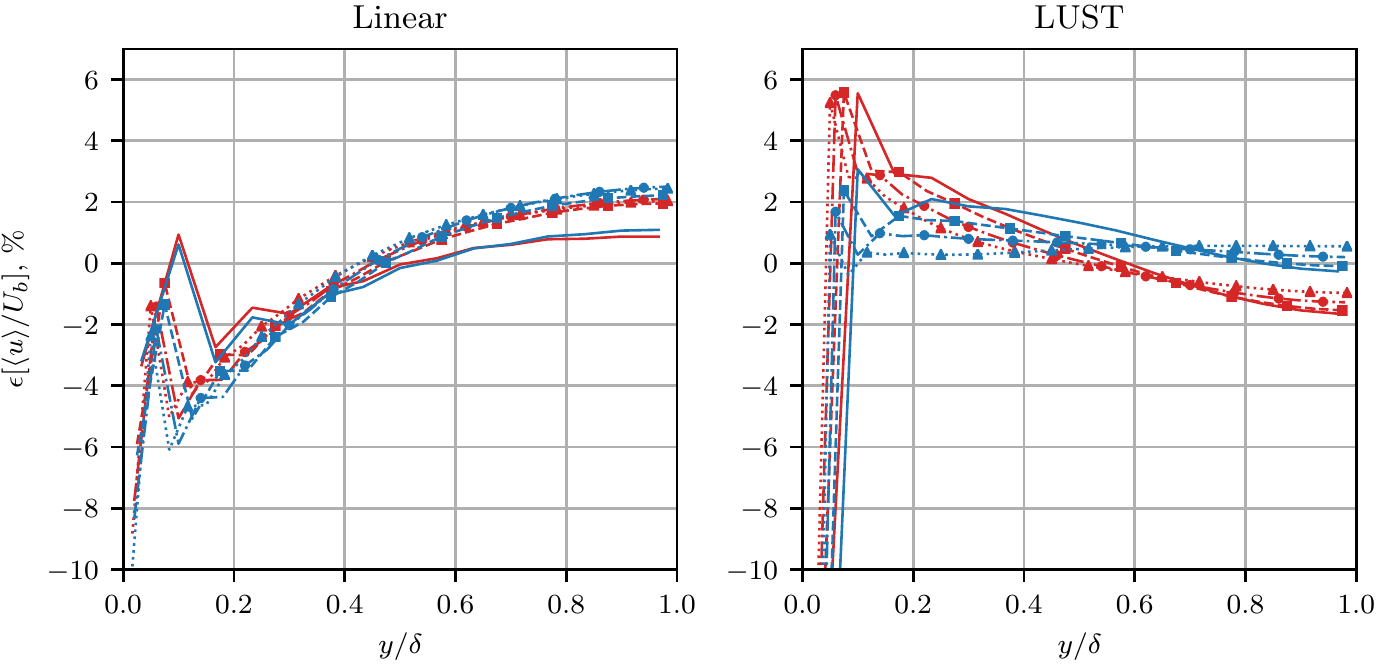}
\caption{
The relative error in the outer-scaled mean velocity profiles from WMLES of channel flow.
Red curves correspond to $h=1^\text{st}$, blue curves to $h=2^\text{nd}$.
A solid line with no markers corresponds to $n/\delta=15$. A dashed line with square markers to $n/\delta=20$. A dashed-dotted line with circle markers to $n/\delta=25$. A dotted line with triangle markers to $n/\delta=30$.}
\label{fig:channel_error_u}
\end{figure}

To fully assess the WMLES it is far from sufficient to only consider the predicted $\mean{\tau_w}$. Attention is now turned to profiles of the obtained flow statistics, starting with the mean of the streamwise velocity, $\mean{u}$.
Figure~\ref{fig:channel_error_u} shows the relative error in the outer-scaled values of $\mean{u}$ as a function of $y/\delta$.
It is observed that the accuracy of all sixteen WMLES with respect to $\mean{u}/U_b$ is acceptable, the relative error not exceeding $2.5\%$ in the core of the channel ($y/\delta > 0.2$).
Closer to the wall, the discrepancies are larger.
The linear scheme produces a non-physical oscillatory solution and the LUST scheme exhibits a very large, 25\%, under-prediction of velocity in the centre of the wall-adjacent cell (point lies outside the axis limit of the plot).
The latter explains the poor performance of the wall model when LUST and $h = 1^\text{st}$ are used, see below.
 
Similarly to $\mean{u_\tau}$, the value of $h$ has a larger effect on the results of the simulations using the LUST scheme. 
An improvement is obtained when $h = 2^\text{nd}$ is used, although it is not as dramatic as in the case of~$\mean{u_\tau}$.
An increase in accuracy with grid refinement is observed only in the case of the LUST scheme.
For the linear scheme, the lowest error overall is, in fact, achieved on the coarsest grid.

\begin{figure}[!htb]
\centering
\includegraphics[]{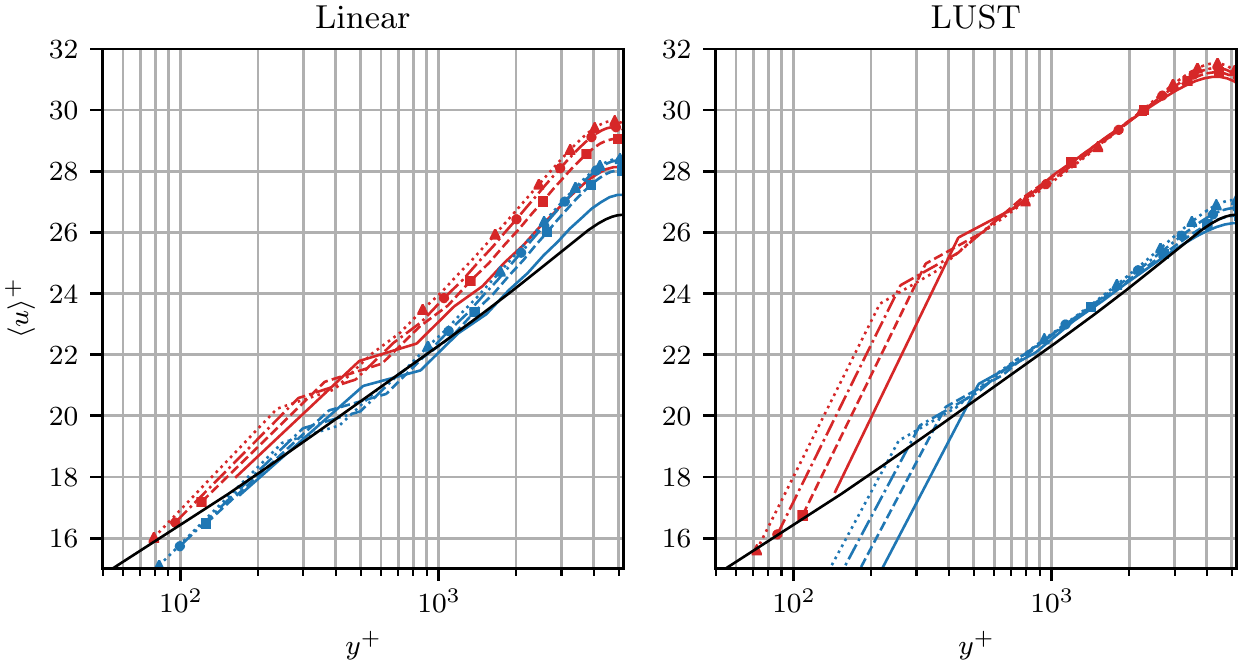}
\caption{Inner-scaled mean velocity profiles from WLES of channel flow.
The black solid line shows the DNS data~\cite{Lee2015}, other line colours and style as in Figure~\ref{fig:channel_error_u}.}
\label{fig:channel_u+_h}
\end{figure}

In Figure~\ref{fig:channel_u+_h}, the obtained mean velocity profiles are shown in inner scaling.
In line with the analysis above, the choice of $h$ mainly manifests itself in the position of the curves along the ordinate.
It is important to note that, by design, the wall model at each time step finds such a $u_\tau$ that would superimpose the point $(h^+, u^+)$ onto the employed law of the wall.
This is clearly seen in the figure, where e.g.~the first data point of all the red curves ($h=1^\text{st}$) lies very close to the DNS profile.
Recall that for the LUST scheme, the value of velocity in the wall-adjacent cell is significantly under-predicted.
This entails that when this cell is used for sampling velocity to the wall model, a corresponding under-prediction in $\olu_\tau$ occurs, in order to shift the value of $u^+$ upwards.
The result is a large log-layer mismatch, clearly seen in Figure~\ref{fig:channel_u+_h}.
Using $h=2^\text{nd}$ provides a remedy because the accuracy of the input velocity signal is much higher.
The explanation for the suboptimal performance is thus the same as the one given in~\cite{Larsson2016}, see Section~\ref{sec:wss_numerical}.  

\begin{figure}[!htb]
\centering
\includegraphics[]{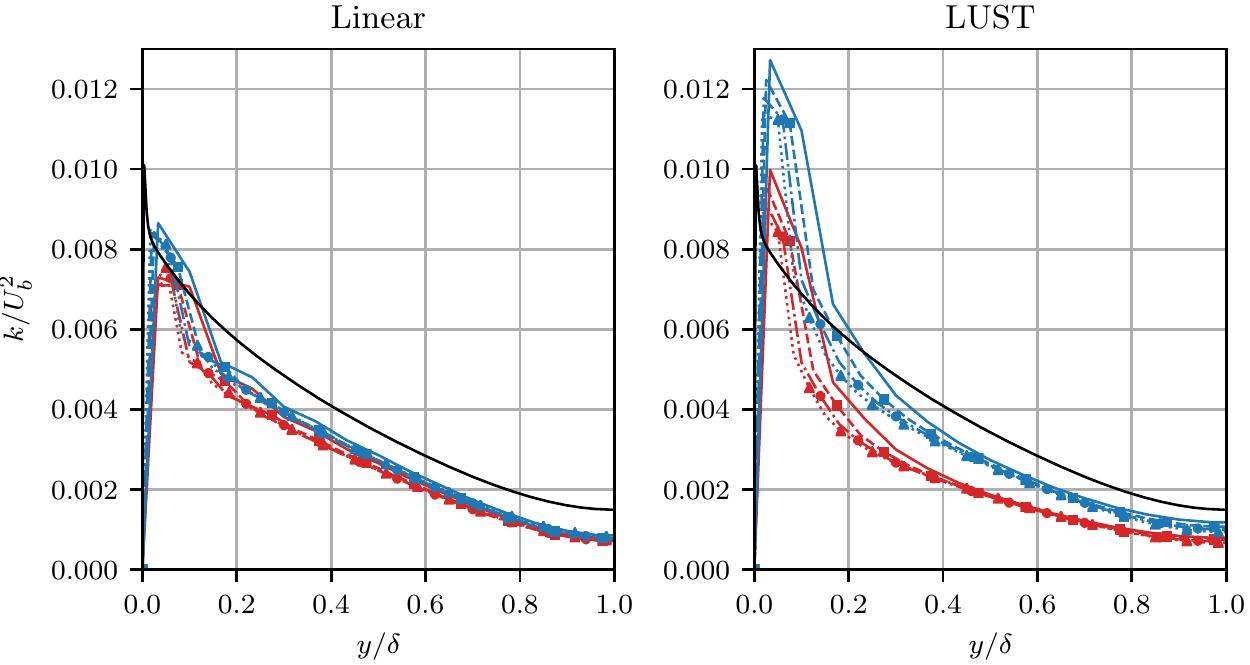}
\caption{
The profiles of the outer-scaled turbulent kinetic energy from WMLES of channel flow. The black solid line shows the DNS data~\cite{Lee2015}, other line colours and style as in Figure~\ref{fig:channel_error_u}.}
\label{fig:channel_k}
\end{figure}

\begin{figure}[!htb]
\centering
\includegraphics[]{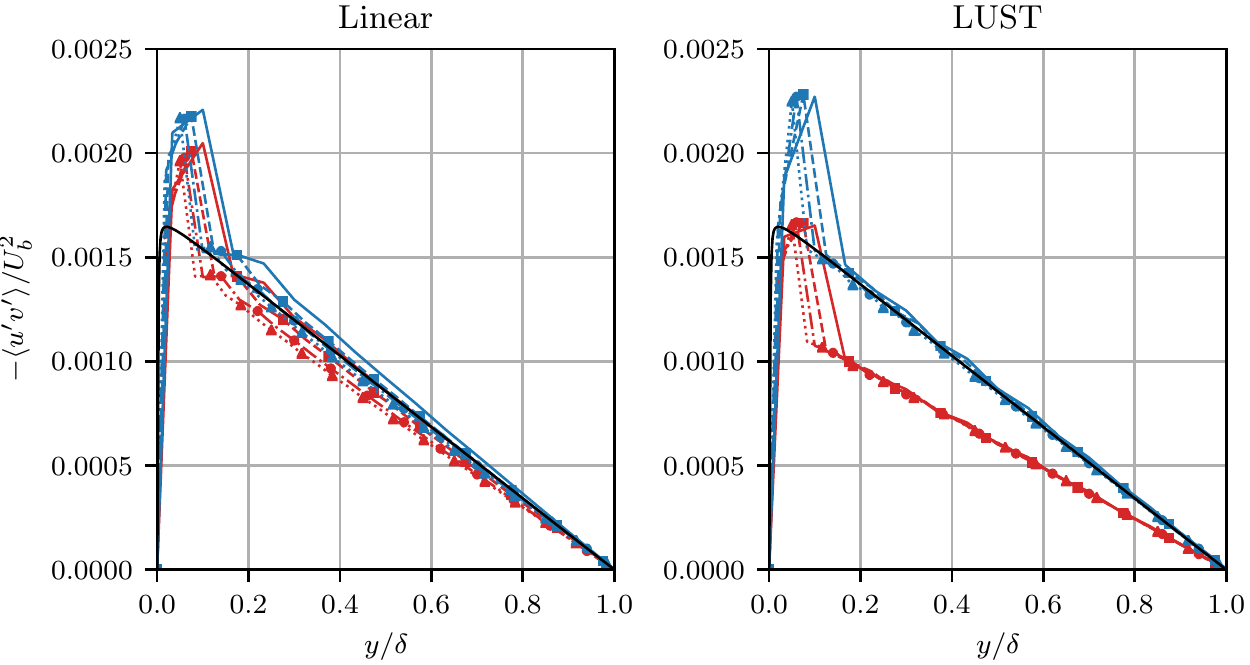}
\caption{
The profiles of the outer-scaled turbulent shear stress from WMLES of channel flow. The black solid line shows the DNS data~\cite{Lee2015}, other line colours and style as in Figure~\ref{fig:channel_error_u}.}
\label{fig:channel_uv}
\end{figure}

Figures~\ref{fig:channel_k} and~\ref{fig:channel_uv} show, respectively, the outer-scaled profiles of the turbulence kinetic energy, $k$, and the turbulent shear stress, $\mean{u'v'}$.
The general trends regarding accuracy are similar to those found for first-order statistics of velocity.
Near the wall, errors are large, whereas in the core of the channel the agreement with DNS is acceptable.
In particular, for $k$, general trends exhibited by LES on coarse meshes are present: over-prediction in the near-wall region and under-prediction in the core of the channel~\cite{Bae2018}.
Similarly to $\mean{u_\tau}$, a dramatic improvement in $\mean{u'v'}$ is observed for the simulations using the LUST scheme when the sampling point is shifted to the second consecutive off-the-wall cell centre.
For LES of channel flow, these two quantities can be shown~\cite{Pope2000} to be connected through the following equation,

\begin{equation}
    \label{eq:channel_momentum}
    \mean{u_\tau}^2 \left(1 - \frac{y}{\delta} \right) = (\nu + \mean{\nu_\text{sgs}}) \frac{du}{dy} - \mean{u'v'}.
\end{equation}
Far from the wall the velocity gradient is not large, which means that getting the correct $\mean{u_\tau}$ leads to accurate values of $\mean{u'v'}$ in that region, and vice versa.

In summary, it can be concluded that all modelling choices have a profound effect on the results of WMLES.
The parameters of the wall model, such as $h$, mainly affect the predicted values of $\mean{u_\tau}$, but their effect on other quantities (considered in outer scaling) is limited.
The most influential modelling parameter overall can be considered to be the choice of the numerical scheme for computing the convective cell-face fluxes.
Based on the obtained results, the LUST scheme is a better choice due to its more consistent behaviour with respect to the resolution of the grid and more accurate prediction of $\mean{u}/U_b$.
It should be noted, however, that the effect of the subgrid-scale model, which was not examined here,  can be anticipated to be at least as influential as that of the interpolation schemes.
The overall accuracy of WMLES is good and on par with what is reported in other studies using similar wall-stress modelling approaches, e.g~\cite{Lee2013, Yang2017}.
In particular, using the LUST scheme, $h=2^\text{nd}$ and $n/\delta =30$ leads to a $1.25\%$ error in $\mean{u_\tau}$ and less than $1\%$ error in $\mean{u}/U_b$ in the core of the channel.

\subsection{Standard vs integrated algebraic wall model formulation}
\label{sec:standard_vs_integrated}
In this section, results from a simulation campaign comparing the performance of the standard and integrated formulation of the algebraic wall model based on Reichardt's law~\eqref{eq:Reichardt} are presented.
All simulations are performed using the $n/\delta = 15$ grid.
This choice was made since the difference in the results obtained using the two formulations can be expected to grow with the wall-normal size of the cell.
Both the linear and the LUST scheme are tested, as well as sampling from the wall-adjacent ($h = 1^\text{st}$) and second consecutive off-the-wall ($h = 2^\text{nd}$) cell.

\begin{table}[htb!]
\caption{Relative error (in percent) in $\langle u_\tau \rangle$ predictions in channel flow simulations using Reichardt's law.}
\centering
\label{tab:channel_utau_integral}
\begin{tabular}{lcccc}
         & \multicolumn{2}{c}{\textbf{Linear}} & \multicolumn{2}{c}{\textbf{LUST}} \\
$\textbf{Formulation}$ & $\bm{h = 1^\text{st}}$                & $\bm{h = 2^\text{nd}}$       & $\bm{h = 1^\text{st}}$               & $\bm{h =2^\text{nd}}$  \\
Standard         & -5.32                           & -1.49                        & -16.62                                & 0.73 \\
Integrated       & -2.26                            & -1.49                         & -13.46                                & 0.80                    
\end{tabular}
\end{table}

The relative error in $\mean{u_\tau}$ is shown in Table~\ref{tab:channel_utau_integral}.
The results obtained using the standard formulation are very similar to those obtained with Spalding's law, see the row corresponding to $n/\delta = 15$ in Table~\ref{tab:channel_utau}.
This is expected, since the difference between the profiles given by these laws is not large, in particular in the logarithmic region.
Using the integrated formulation improves the results by $\approx 3$ percentage points, when sampling from the wall-adjacent cell is used, corroborating the results in~\cite{Temmerman2003}.

In the case of $h =2^\text{nd}$, the accuracy is not improved.
This can be explained by the fact that the wall-normal variation in $\mean{u}^+$ is highest near the wall, leading to a significant difference between the point-wise value of $\mean{u}^+(h^+)$ and the corresponding wall-normal average of  $\mean{u}^+$ across the extent of the WMLES cell.
Farther from the wall, the velocity profile varies slower with $y$,  and both formulations yield quite similar results.

\section{Application to flow over a backward-facing step}
\label{sec:bfs}

The developed library was also used for WMLES of the more complicated case of flow over a backward-facing step (BFS).
In this section, a discussion of the case set-up is first followed by a general overview of the flow and its features.
Then results from a simulation campaign similar to the one reported in Section~\ref{sec:channel_parameter_study} for channel flow are discussed, analysing the influence of several important modelling choices. 
Finally, the performance of the ODE-based wall models included in the library is investigated.

\subsection{Case set-up}
\label{sec:bfs_case_setup}

\begin{figure}[!htb]
  \centering
\includegraphics[height=2.6in]{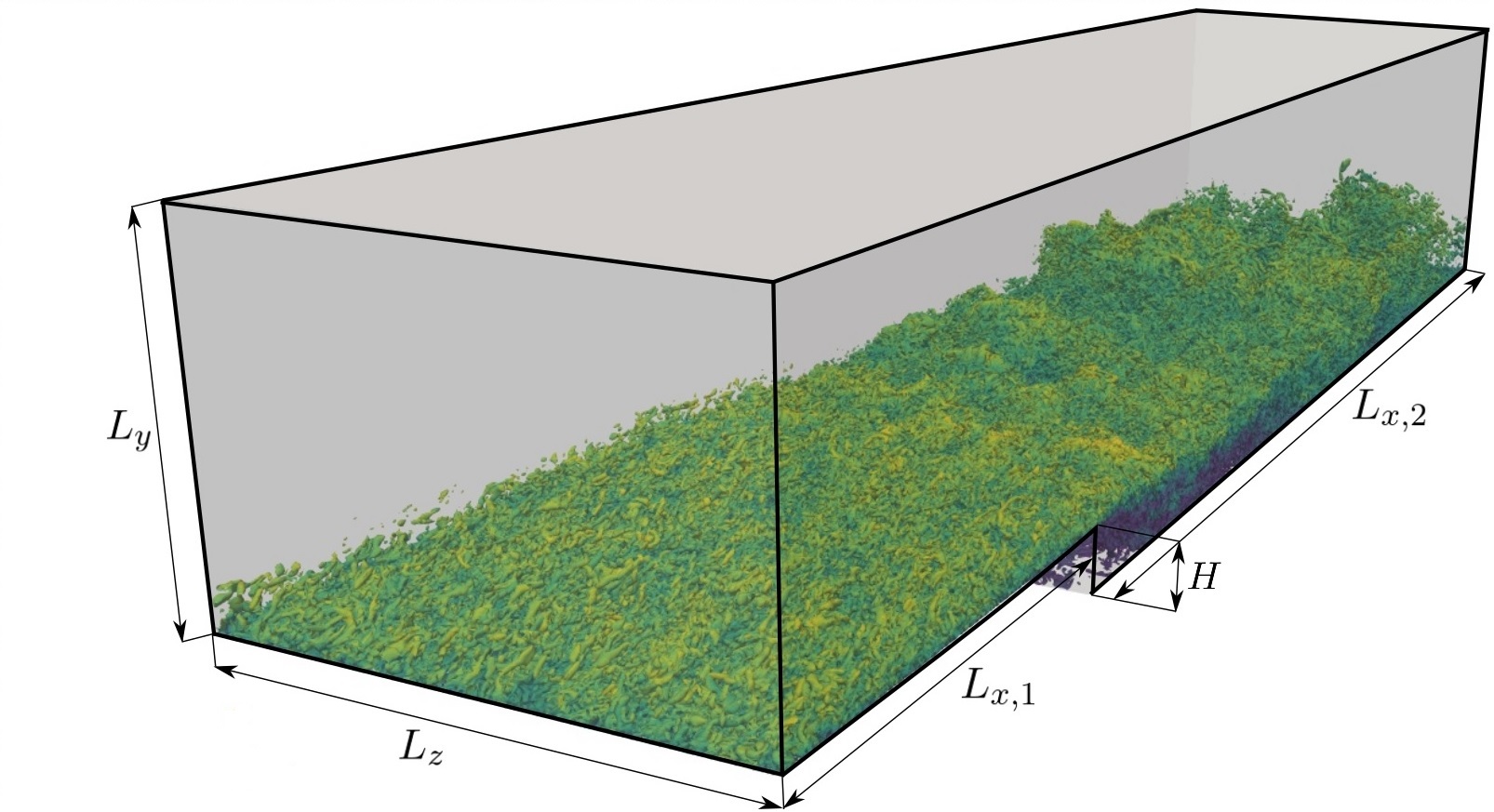}
  \caption{The computational domain of the BFS simulation. Resolved turbulent structures are visualised by iso-surfaces of $\lambda_2$-criterion, coloured by instantaneous streamwise velocity}
    \label{fig:domain}
\end{figure}

Figure~\ref{fig:domain} shows the computational domain along with resolved turbulent structures visualised using the~$\lambda_2$-criterion.
The flow over a BFS can be roughly divided into three regions.
\textit{i)}~The turbulent boundary layer, entering the domain at the inlet and developing along a flat-plate located upstream of the step.
\textit{ii)}~The shear layer formed by the detached TBL, and the recirculation zone beneath it.
\textit{iii)}~The recovery region, following the reattachment of the boundary layer.

The parameters fully defining the flow are the Reynolds number of the separating TBL, and $\rey_{H}$, i.e. the Reynolds number based on the step-height, $H$ and the free-stream velocity $U_0$.
The simulations were set up to match the experiment of Jovic~\cite{Jovic1996}, in which $\rey_H = 37\, 000$.
For the separating TBL, a reference value for the momentum thickness-based Reynolds number is provided, $\rey_\theta \approx 3\,600$.
Also, the ratio $\delta_{99}/H = 0.8$ is given, where $\delta_{99}$ is defined as the wall-normal distance at which $\mean u = 0.99U_0$.
Both the momentum thickness, $\theta$, and $\delta_{99}$ are measured at a station located $x/H = -1.05$, where $x=0$ is located at the step.

In order to introduce turbulence at the inflow, a precursor turbulent channel flow simulation is used according to the method proposed in~\cite{Mukha2017}.
Instantaneous velocity values are sampled in the course of the precursor simulation, from a plane normal to the streamwise direction and spanning the interval $[0, \delta]$ in the wall-normal direction. 
These velocity values are then prescribed at the inlet of the BFS domain without any further manipulation.
Above $y=\delta$, the free-stream velocity $U_0$ is prescribed, where $y=0$ corresponds to the location of the flat plate upstream of the step.
Hence, the precursor channel flow must be set up in such a way that the mean integral characteristics of the sampled velocity fields match those desired for the inlet TBL, here $\rey_\theta$ and $\delta_{99}/H$.
In a more applied setting, the dimensional values of the components forming the latter two quantities would also be fixed, and the dimensional characteristics of the channel flow ($\delta$,~$U_b$,~$\nu$) would have to be chosen accordingly.
Here, however, it is possible to define the parameters of the precursor first, and let that drive the set-up the BFS, based on the values of $\rey_H$, $\rey_\theta$ and $\delta_{99}/H$.

In the case of a WRLES, it would be possible to set up the precursor to match both $\rey_\theta$ and $\delta_{99}/H$ simultaneously. 
But for WMLES this turns out to be difficult due to the level of accuracy of the mean velocity profile, which was shown to vary significantly on the modelling choices, see Section~\ref{sec:channel_parameter_study}.
In particular, a reliable estimate of $\rey_\theta$ is difficult to obtain, because the momentum thickness is computed using the whole mean velocity profile, including the inaccurate solution in the inner layer.
The value of $\delta_{99}$ was generally observed to be more robust and could be quite accurately estimated as $\approx 0.85\delta$.
As discussed above, the value of $\delta$ can be chosen freely.
Here, \mbox{$\delta = 1$ m} is adopted for simplicity, leading to \mbox{$\delta_{99} \approx 0.85$ m} at the inlet.
However, to determine $H$ from the given ratio of $\delta_{99}/H$, the value of $\delta_{99}$ at $x/H = -1.05$ is required.
It was observed that while $\delta_{99}$ initially grows in the streamwise direction, the favourable pressure gradient present immediately upstream of the step mitigates this growth, thus allowing to use \mbox{$\delta_{99} \approx 0.85$} m as a reliable prediction of the quantity's value at $x/H = -1.05$.
This leads to $H = 1.0625$ m.
Although the above computations are based on several approximations, the error in the values of $\delta_{99}/H$ obtained in the simulations did not exceed 10\%.

The free-stream velocity $U_0$ is matched to the center-line velocity in the precursor channel flow.
The latter can be robustly predicted to be $\approx 1.12$~m/s, given the chosen value of 1~m/s for the bulk velocity $U_b$. 
Having defined $H$ and $U_0$, $Re_H = 37\, 000$ is obtained with the following value of the kinematic viscosity, \mbox{$\nu = 3.1875\cdot 10^{-5}$ $\text{m}^2/\text{s}$}.
The physical parameters of both the BFS and the precursor channel flow are thus fully defined.

For the BFS simulation, four more geometrical parameters have to be specified.
The first is the length of the flat plate upstream of the step, $L_{x, 1}$.
The value has to be sufficiently large in order for the errors associated with the prescribed inflow to become negligible.
Here the value of $L_{x, 1} = 8H$ was chosen, which corresponds to $10 \delta_{99}$, measured at the inlet.
The length of the downstream plate $L_{x, 2}$ has to allow for an analysis of the recovery of the TBL following reattachment.
Here, $L_{x, 2} \approx 32H$ is used.
Based on the  expansion ratio $(L_y + H)/L_y = 1.19$ in the reference experiment~\cite{Jovic1996}, the height $L_y = 5.26H$ was chosen. 
Finally, width $L_z = 8\delta  \approx7.53H$ was chosen, which is large enough to avoid spurious periodicity effects.

Apart from the boundary condition at the inlet, which was discussed above, the following conditions are used at the other boundaries.
At the flat plates, wall modelling is applied.
At the outlet, the pressure is set to zero and a homogeneous Neumann condition for velocity values is used.
The top boundary is treated as a symmetry plane, matching the set-up of~\cite{Jovic1996}.
Finally, a periodic condition is applied to the sides of the domain.

The part of the domain occupied by turbulent flow is meshed using cubic cells of equal size.
Thus, as in the case of channel flow, the resolution of the mesh is fully specified by the value of $n/\delta$.
In the region occupied by the free stream, the size of the cells in the wall-normal direction is rapidly increased.
A grid with matching resolution is used in the precursor channel flow.
All simulations were first run for $\approx  8T_{ft}$, where $T_{ft} = 37.8$ s is the domain flow-through time.
Afterwards, time-averaging was started and continued for another $\approx 45T_{ft}$.
The time-step size used was $\Delta t = 0.01$ s, which corresponds to $\approx 0.01 H/U_0$.
The employed wall modelling, grid density, and convective cell-flux interpolation scheme are separately discussed for each of the simulations presented below. 

\subsection{Overview of the flow}
This section gives a general overview of the features of the flow over a BFS.
This discussion is supported by plots of results obtained from a particular simulation that uses the combination of modelling parameters, which was found to give the best predictive accuracy, based on the study presented in Section~\ref{sec:bfs_parameter_study} below.
A mesh with resolution $n/\delta = 30$ is employed, and to decrease the overall number of cells to $\approx 20.5\cdot 10^6$, the length of the downstream flat plate is in this particular simulation lowered to $L_{x,2} \approx 24.5H$.
The LUST scheme is used for convective cell-face flux interpolation.
The algebraic model based on Spalding's law~\eqref{eq:spalding} is used for wall modelling, with the following distribution of $h$ over the downstream flat plate: $h = 1^\text{st}$, for $x/H<8$; $h=2^\text{nd}$, for  $x/H>8$.
At the flat plate upstream of the step, $h=2^\text{nd}$ is used.

\begin{figure}[!htb]
\centering
\includegraphics[width=\linewidth]{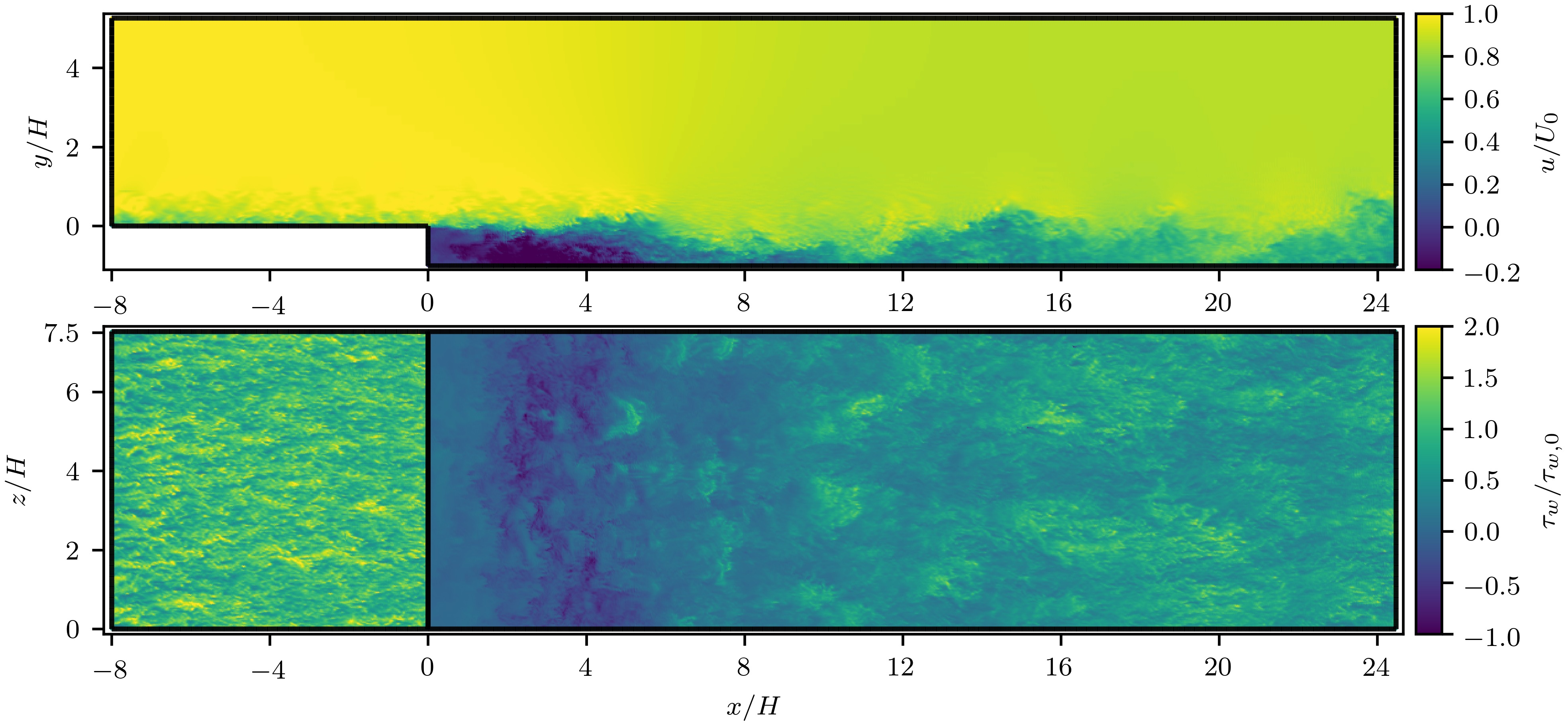}
\caption{\textit{Top:} A snapshot of the scaled instantaneous streamwise velocity, $u/U_0$. \textit{Bottom:} The normalised instantaneous wall shear stress, $\tau_w/\tau_{w,0}$, where $\tau_{w,0}$ is the average wall shear stress at the inlet of the domain.}
\label{fig:tau_bfs}
\end{figure}

Figure~\ref{fig:tau_bfs} shows a snapshot of the distribution of $u$ over an $x$-$y$ cut-plane of the domain and the distribution of $\tau_w$ over the walls (as predicted by the wall model).
Upstream of the step, the turbulent structures in the boundary layer are seen, and the patterns of the predicted $\tau_w$ are similar to those obtained for channel flow, see Figure~\ref{fig:linear_lust}.
Downstream, the separated TBL forms a shear layer.
It is evident from the figure, that the shear layer has a large effect on the flow throughout the whole extent of the domain downstream of the step.
Both in the recirculation zone, and after reattachment, the distribution of $\tau_w$ does not resemble that observed upstream of the step.
This indicates that this flow presents a challenge for wall models based on a law of the wall, such as the one used here.

\begin{figure}[!htb]
\centering
\includegraphics[]{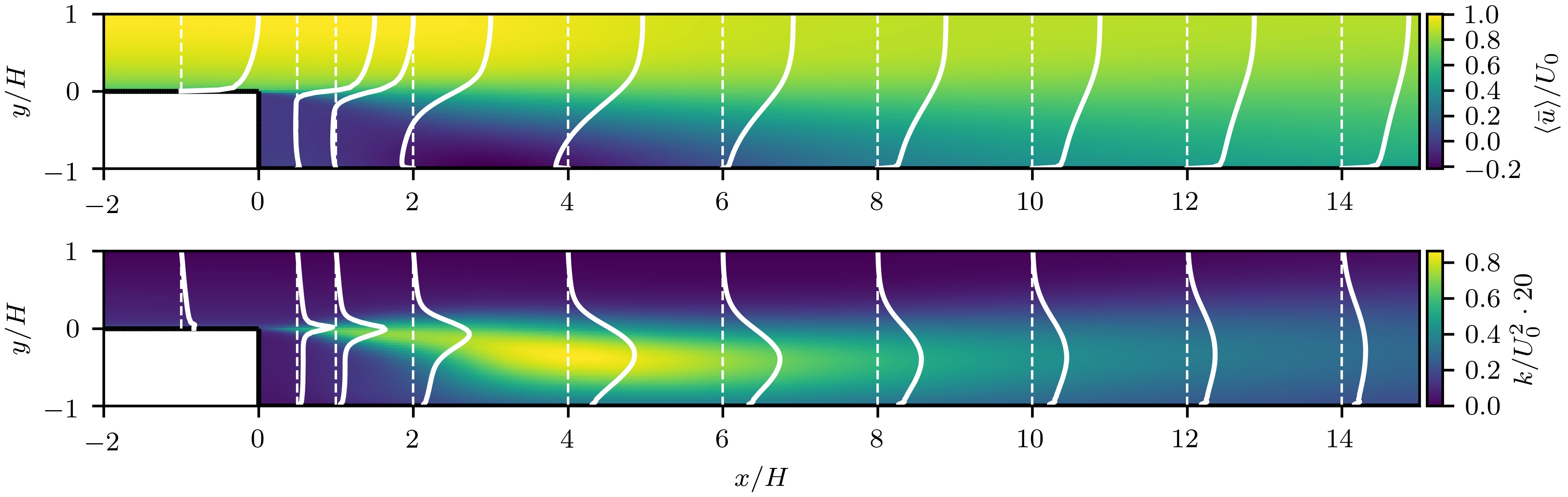}
\caption{
\textit{Top:} Distribution of the scaled mean streamwise velocity, $\mean u/U_0$.
\textit{Bottom:} Distribution of the scaled turbulent kinetic energy, $20k/U_0^2$. Profiles of both quantities at selected stations are shown with solid white lines.
The zero-level for each profile is indicated by a dashed white line.}
\label{fig:bfs_overview}
\end{figure}

The top and bottom plots in Figure~\ref{fig:bfs_overview} show the distribution of the mean streamwise velocity and the resolved turbulent kinetic energy, respectively.
The profiles of these quantities at selected stations are also shown.
It is seen that up to $x/H = 1$ the velocity values in the recirculation region are low, as well as the level of turbulent fluctuations.
Downstream the flow is affected by the detached shear layer, with the highest values of~$k$ observed at $x/H \approx 4$, which is~$\approx 1.55H$ upstream of the mean reattachment point, $x_r \approx 5.55H$.
The influence of the turbulent shear layer is present all the way down to the outlet, with the peak in $k$ observed at~$y/H \approx -0.5H$ significantly exceeding  in magnitude the near-wall peak associated with the recovering TBL.
This is also reflected in the mean velocity profiles, which clearly differ from those of a canonical zero-pressure-gradient flat-plate TBL.

\begin{figure}[!htb]
\centering
\includegraphics[clip=true]{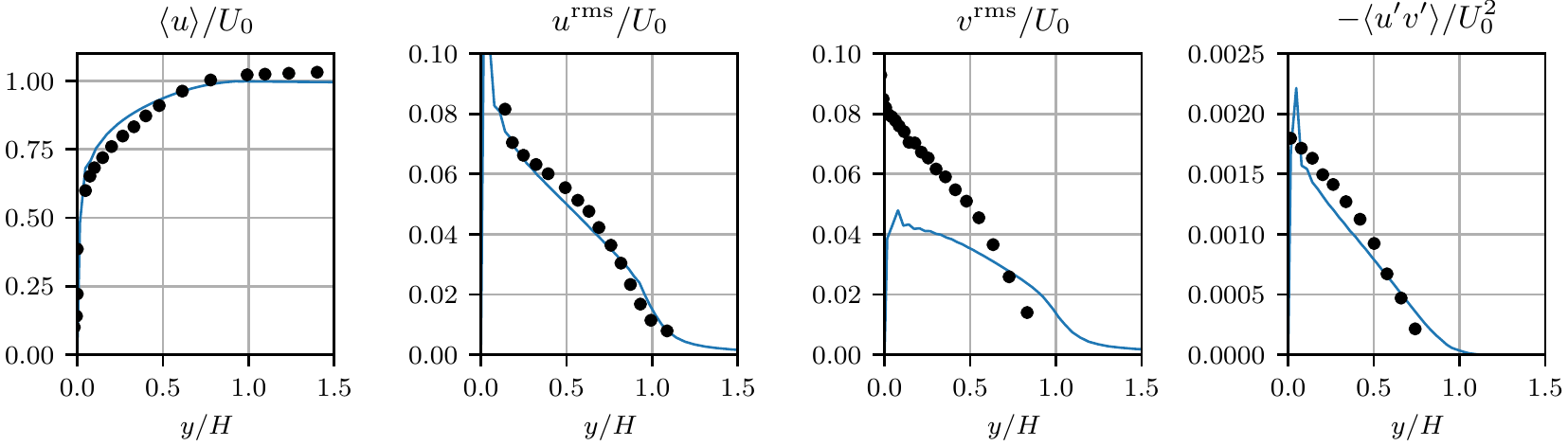}
\caption{Profiles of $\mean u /U_0$, $u^\text{rms}/U_0$, $v^\text{rms}/U_0$, and $-\mean{u'v'}/U^2_0$ taken at $x/H = -1.05$.
Blue lines show the WMLES results, black circles show the measurements reported in~\cite{Jovic1996}.}
\label{fig:bfs_inflow}
\end{figure}

After the above overview of the flow, the discussion is now focused on a more detailed description of certain features of the flow.
Firstly, the profiles of the mean streamwise velocity and three components of the Reynolds stress tensor at $x/H = -1.05$ are considered, see Figure~\ref{fig:bfs_inflow}.
Recall that this is the only location upstream of the step where reference experimental data are available~\cite{Jovic1996}.
In particular, the value of $\delta_{99}/H = 0.8$, measured at this station, was used to define the inflow TBL.
The value of this quantity obtained in the simulation is  $\approx 0.79$.
This, in conjunction with the good agreement for $\mean u / U_0$ observed in the figure, allows to conclude that the simulation is successful at reproducing the set-up of the reference experiment.
Based on the results for channel flow, a lower level of agreement can be expected for second-order statistical moments (see Figure~\ref{fig:channel_k}).
Generally, for LES on coarse meshes over-prediction of $u^\text{rms}$ and under-prediction of $v^\text{rms}$ and $w^\text{rms}$ is a commonly occuring error pattern~\cite{Bae2018}.
Here, a remarkably good agreement with the experiment is found for $u^\text{rms}$, whereas the predicted $v^\text{rms}$ values are indeed lower than those of the reference.
For the turbulent shear stress the agreement with the experiment is good, although, as in the case of channel flow (see Figure~\ref{fig:channel_uv}), a non-physically large peak is observed near the wall.

\begin{figure}[!htb]
\centering
\includegraphics[]{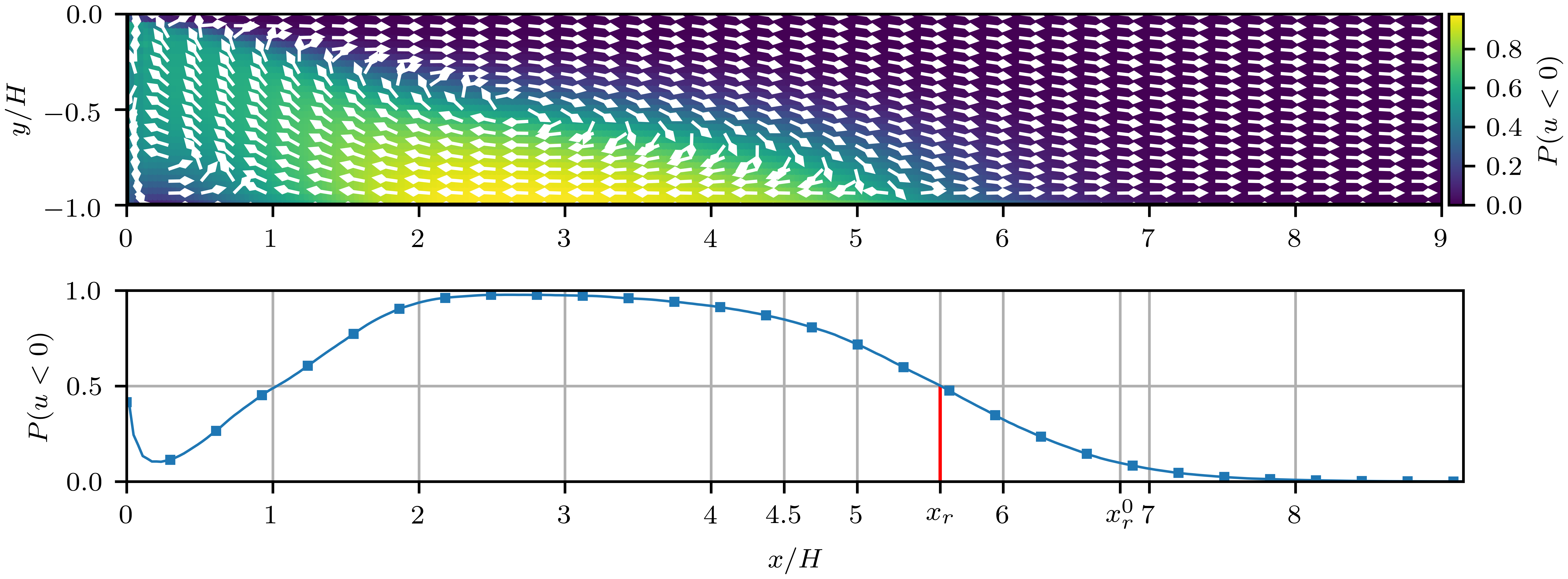}
\caption{\textit{Top:} The distribution of the probability of back-flow in the recirculation bubble.
The direction of the mean flow is shown with white arrows.
\textit{Bottom:} The probability of back-flow in the wall-adjacent cells as a function of $x/H$.
 The location of the reattachment point (computed as the location where $P(u < 0) = 0.5$) is indicated with a red line. }
\label{fig:bfs_backflow_accurate}
\end{figure}

Attention is now turned to the recirculation region.
The direction of the mean flow is shown in the bottom plot of Figure~\ref{fig:bfs_backflow_accurate}.
Besides for the main separation bubble, a small secondary bubble is present in the lower corner of the step.
The same plot also shows the distribution of the probability of back-flow, $P(u < 0)$.
This quantity is computed by time-averaging the $(1 - \text{sgn}(u))/2$ field in the course of the simulation, where $\text{sgn}$ is the sign function.
It is observed that the direction of the flow is highly intermittent, with back-flow predominant only in the interval $x/H \in [2, 4]$.
In the bottom plot of the same figure, $P(u < 0)$ in the wall-adjacent layer of cells is shown.
Up to $x/H \approx 1$ forward-flow dominates due to the secondary bubble.
The probability of back-flow then continues to increase, reaching values close to 1, but starts to decline later at $x/H \approx 3$.
The mean reattachment point, $x_r/H \approx 5.55$, can be computed as the location where back- and forward-flow are equally probable, see the red line in the plot.
This obtained value deviates quite significantly from the value of $x^0_r/H \approx 6.8$ measured in the reference experiment ~\cite{Jovic1996}.
It will be shown below that this quantity is highly sensitive to the modelling parameters of the WMLES.
Also, evidence towards low accuracy in the experimentally obtained value will be given.
The probability of back-flow becomes essentially zero at $x/H \approx 8$.
This is the reason why this location is chosen for switching the wall model to sampling from the second consecutive off-the-wall cell.

\begin{figure}[!htb]
\centering
\includegraphics[clip=true]{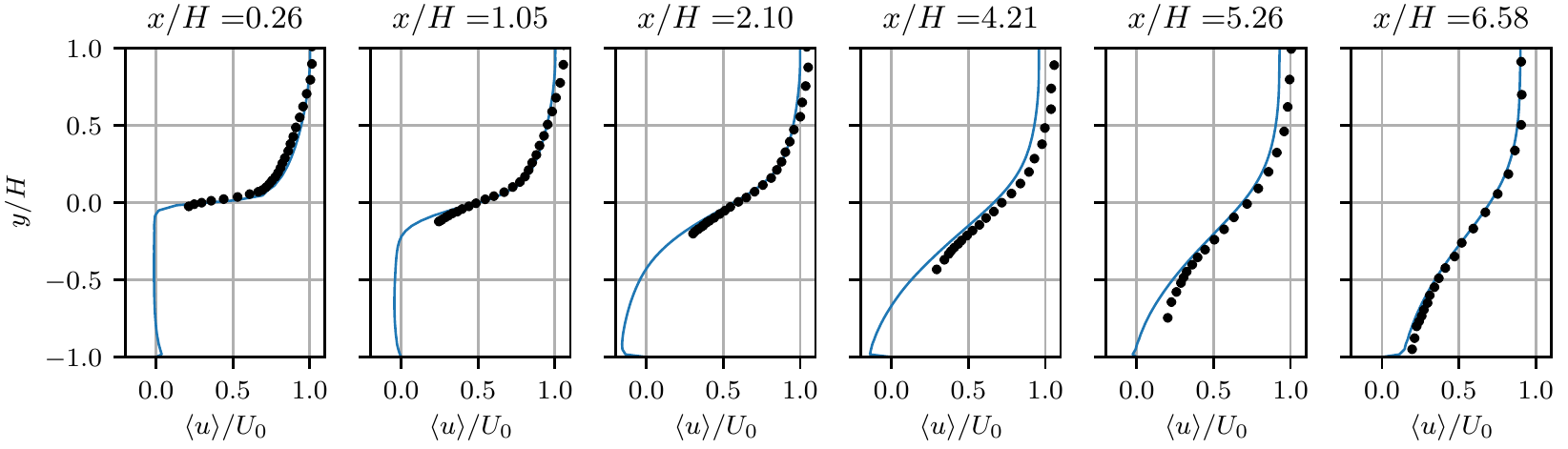}
\caption{Mean streamwise velocity profiles in the detached shear layer and recirculatoin zone.
Blue lines show WMLES results, and black circles the experimental data from~\cite{Jovic1996}.
}
\label{fig:bfs_u_shear_accurate}
\end{figure}

The mean velocity profiles obtained in the shear layer and the recirculation zone below it are shown in Figure~\ref{fig:bfs_u_shear_accurate}.
The agreement with the reference experimental data is very good at all five stations, although some deviation is observed at $x/H = 5.26$, below $y/H = -0.5$.
Remarkably, no back-flow is present in the experimental profile at $x/H = 6.58$ in spite of the reported mean reattachment point being $x^0_r/H \approx 6.8$.
In fact, even at $x/H = 5.26$ it is reasonable to assume that no back-flow is registered based on the shape of the profile, even though data below $y/H = -0.75$ is not provided.
Thus, it appears that the reported value of $x^0_r$ is over-predicted, and the agreement between the simulation and the experiment with respect to this quantity may actually be better.
Note also that the thin boundary layer under the recirculation bubbles is not properly resolved by the grid, with at most three points located between the wall and the wall-normal location of the maximum back-flow.

\begin{figure}[!htb]
\centering
\includegraphics[clip=true]{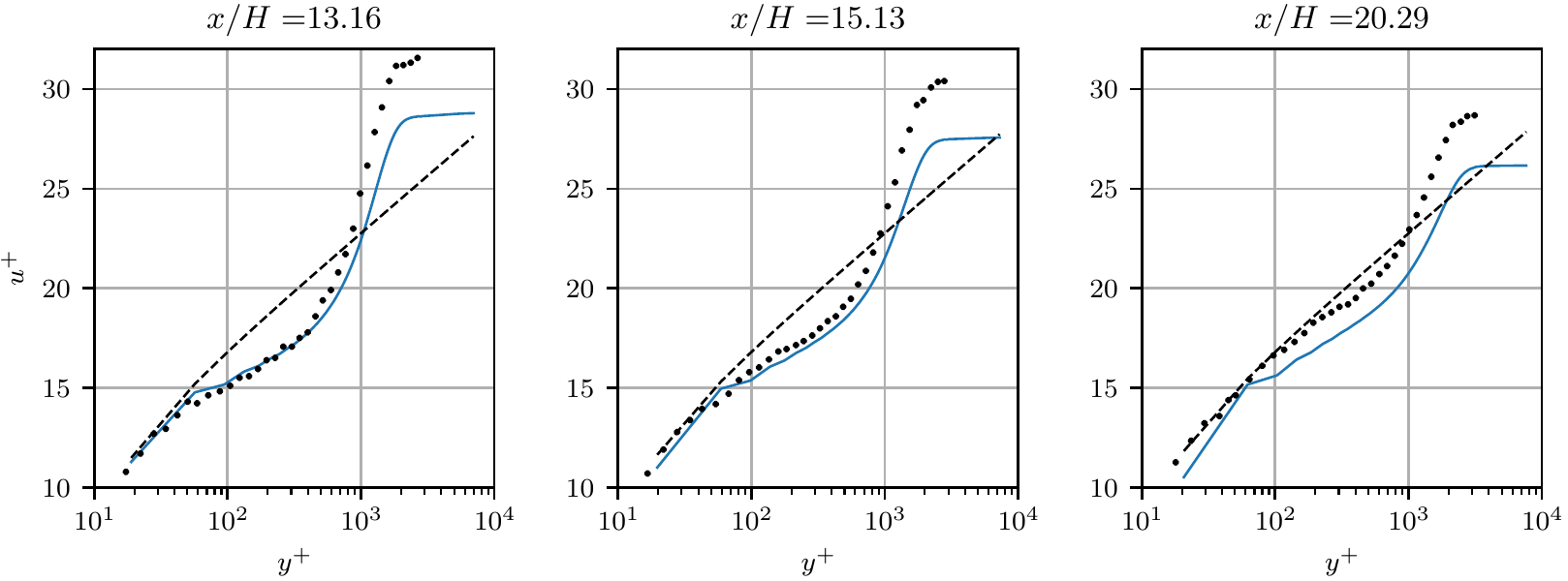}
\caption{The inner-scaled mean streamwise velocity profiles at three stations downstream of the mean reattachment point.
Blue lines show WMLES results, black circles the experimental data from~\cite{Jovic1996}, and dashed black lines show Spalding's law~\eqref{eq:spalding}.}
\label{fig:bfs_u_recovery_accurate}
\end{figure}

The discussion continues with the analysis of the recovering TBL.
Figure~\ref{fig:bfs_u_recovery_accurate} shows the inner-scaled mean streamwise velocity profiles at three downstream locations.
The recovery of the log-law can be seen, and at $x/H = 20.29$ a good agreement with Splading's law is found in the inner and log-law regions.
At the two stations upstream, however, agreement with the law is found only at $y^+ < 60$.
Falling into this region are the locations of the sampling points at all three stations, which are found between  $y^+\approx 55$ and $\approx 65$.
This leads to accurate predictions of the mean wall shear stress, see below.
In the outer layer, the recovery process is much slower, and the profiles exhibit a shape typical of TBLs under a strong adverse pressure gradient.
However, it will be shown that the strength of the pressure gradient in this region is, in fact, negligible.
The profile shape thus has to be attributed to the influence of the shear layer, as concluded by Jovic~\cite{Jovic1996}.
The agreement between the WMLES and the experimental data is good, in particular in the inner region.
Some discrepancy is present in the outer region, however, with the experimental data exhibiting a steeper wall-normal gradient.
This may indicate that a higher grid resolution is needed to properly resolve the interaction between the shear layer and the recovering TBL. 

\begin{figure}[!htb]
\centering
\includegraphics[clip=true]{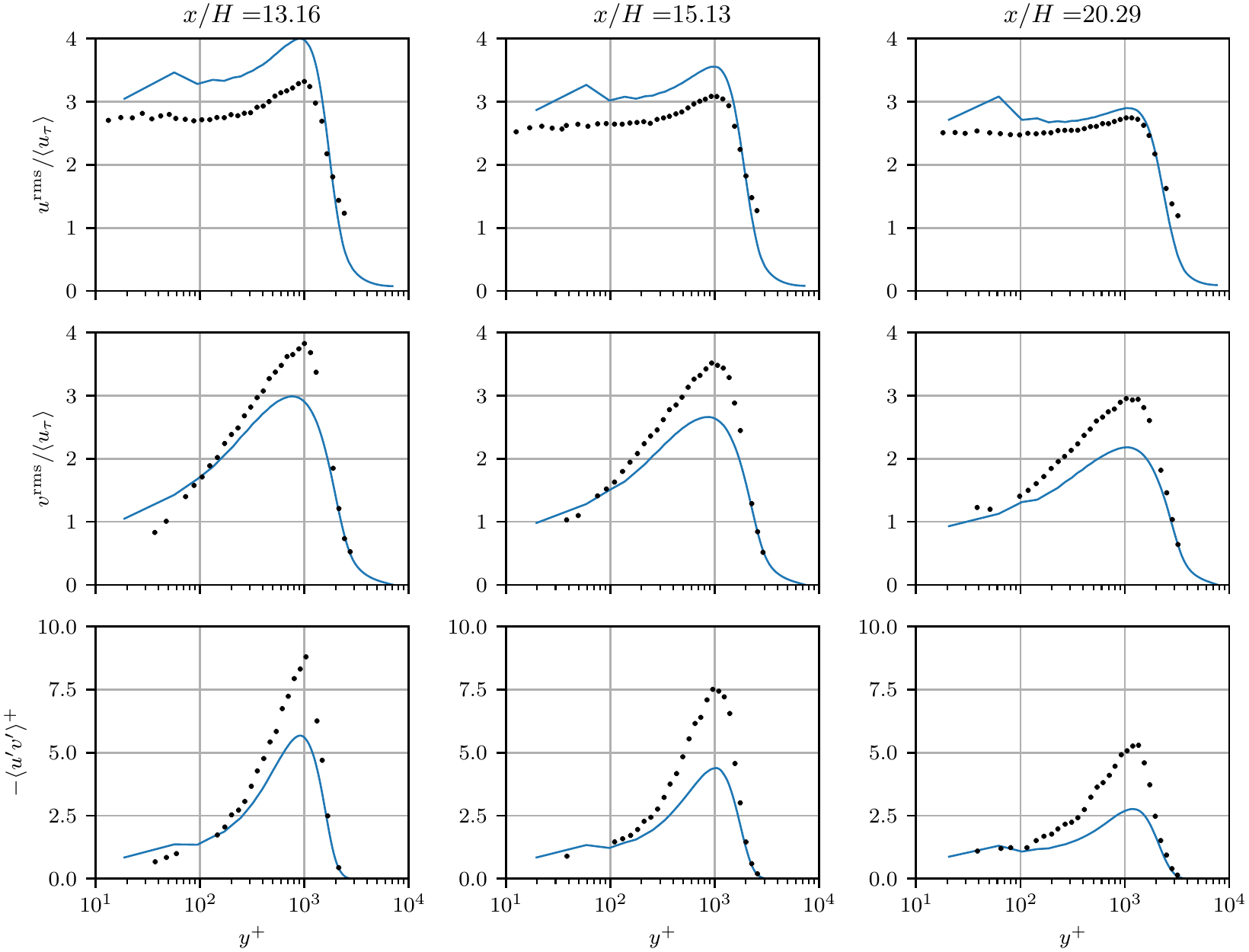}
\caption{The inner-scaled profiles of $u^\text{rms}$, $v^\text{rms}$ and $-\mean{u'v'}$ of the recovering TBL taken at three downstream locations.
Each column of plots corresponds to an axial station and each row of plots corresponds to a Reynolds stress component.
Blue lines show WMLES results, and black circles the experimental data from~\cite{Jovic1996}.}
\label{fig:bfs_re_recovery_accurate}
\end{figure}

Three components  of the Reynolds stress tensor at the same downstream locations are shown in Figure~\ref{fig:bfs_re_recovery_accurate}.
For all the quantities, a large peak is present at $y^+ \approx 1000$, which can be attributed to the detached shear layer.
The magnitude of the peak decays with $x$, but at a very slow pace, see also bottom plot in Figure~\ref{fig:bfs_overview}.
This behaviour has been reported for other separating flows as well, see e.g.~\cite{Song2000}.
In contrast, no near-wall peak typical of a canonical TBL is present.
The agreement between the WMLES and the reference is overall acceptable, in particular for $u^\text{rms}$.
The WMLES values of $v^\text{rms}$ and $\mean{u'v'}$ away from the wall are, however, lower than the corresponding experimental values.
Recall that for the former quantity the same discrepancy was observed at a station upstream of the step and also for channel flow simulations.

\begin{figure}[!htb]
\centering
\includegraphics[clip=true]{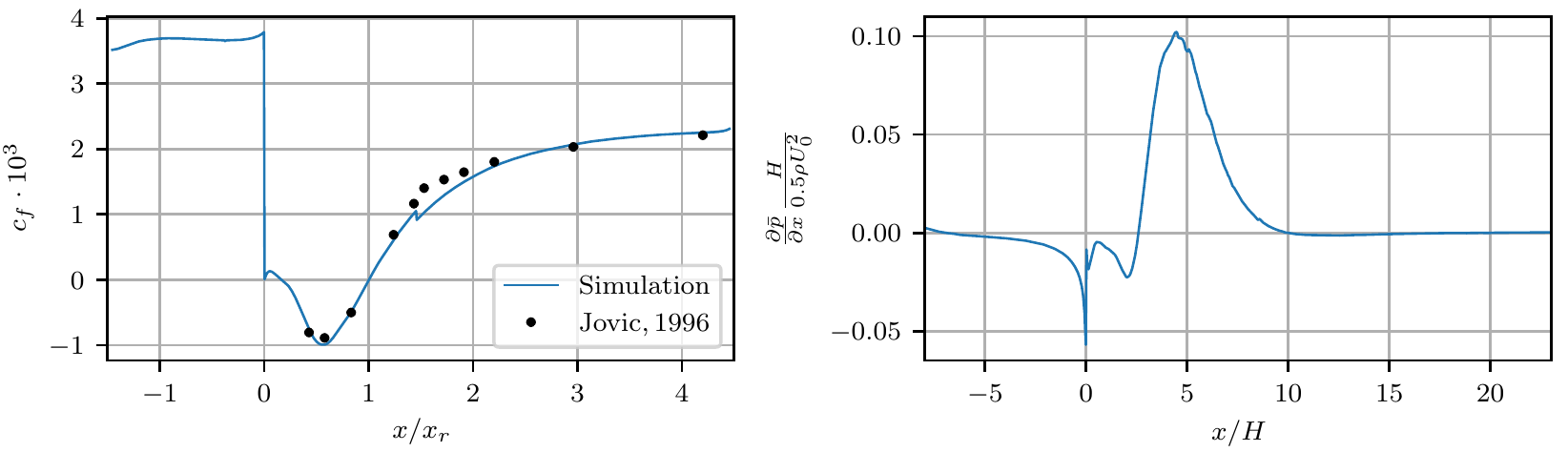}
\caption{\textit{Left:} Skin friction coefficient, $c_f$, on the bottom wall.
\textit{Right:} Normalised streamwise pressure gradient on the bottom wall.}
\label{fig:bfs_cf_accurate}
\end{figure}

In the last part of this section, quantities directly connected to wall modelling are considered.
The left plot in Figure~\ref{fig:bfs_cf_accurate} shows the distribution of the skin friction coefficient, $c_f = \mean{\tau_w}/(0.5\rho U_0^2)$, exhibiting the wall models ability to correctly predict the mean wall shear stress.
The abscissa is scaled by $x_r$, thus concealing any discrepancies in the prediction of this quantity between the WMLES and the experimental data.
This is motivated by the uncertainty regarding the accuracy of the prediction of $x_r$ in the latter.
Directly downstream of the inflow, a transition period manifested in a rise of $c_f$ is present due to the inflow boundary condition.
Further on, $c_f$ slightly decreases with $x$, thus behaving similarly to a flat plate zero-pressure-gradient TBL.
Prior to separation, the effect of the favourable pressure gradient becomes significant enough to affect $c_f$, resulting in its increase.
Directly downstream of the step, the $c_f$-values experience a slight bump associated with the secondary recirculation bubble, which is followed by negative values in the region occupied by the main recirculation bubble.
The agreement with the experimental data is remarkably good here, with almost no error in the value of the negative peak.
This indicates that simple algebraic wall models are at least in some cases capable of producing accurate $\mean{\tau_w}$ predictions even in regions where the underlying law of the wall is not valid.
A small abrupt drop in~$c_f$ is visible at the downstream location corresponding to $x/H = 8$, which is the point where the model switches to sampling from the second off-the-wall cell.
Downstream of this location the prediction of $c_f$ is also good, which can be attributed to the fact that the sampling point is located in the region where Spalding's law overlaps with the mean velocity profile.

In the right plot of Figure~\ref{fig:bfs_cf_accurate}, the distribution of the streamwise pressure gradient on the horizontal walls is shown.
As it was mentioned above, a strong favourable pressure gradient is present directly upstream of the step.
Downstream, after a short transition period, the pressure gradient changes sign and increases in magnitude, peaking at a location $\approx H^2$ upstream of $x_r$.
This increase is directly associated with the expansion of the flow domain in the wall-normal direction.
Note that the direction of the flow near the wall in this region is  reversed so with respect to that the pressure gradient is, again, favourable.
Further downstream the magnitude of the pressure gradient decreases and for $x/H > 10$ it can be considered negligible.
The region where the pressure gradient is adverse with respect to the near-wall flow is thus confined to $[x_r, 10H]$.

In conclusion, it is considered that the obtained agreement between the WMLES and the experimental data is good for both the mean velocity profiles and for the skin fiction coefficient.
The prediction of the latter in the recirculation zone is, in fact, surprisingly accurate given the fact that a simple algebraic wall model is employed.
Some discrepancy in the form of the mean velocity profiles has been observed in the recovering TBL, which may indicate that a higher grid resolution is needed to capture the interaction between the TBL and the detached shear layer.
Similarly to the results of the channel flow simulations, second-order statistics are computed with less precision, but the overall shape of the profiles is correct.

\subsection{Influence of $n/\delta$, $h$, and interpolation scheme for convective fluxes}
\label{sec:bfs_parameter_study}

Here, a study similar to that reported for turbulent channel flow in Section~\ref{sec:channel_parameter_study} is presented.
The goal is again to analyse the influence of the same three modelling parameters on the predictive accuracy of the WMLES, i.e.~that of the grid resolution $n/\delta$, the distance to the sampling point $h$, and the interpolation scheme for the convective cell-face fluxes.
It is interesting to see whether the conclusions of the channel flow study will remain valid for the more complicated case of the flow over a BFS.
Due to limitations in computational resources, only two values of $n/\delta$ are considered here, 15 and 20.
Two values of $h$ for the downstream wall are considered, $h = 1^\text{st}$ and $h = 2^\text{nd}$.
Based on the results of the channel flow simulations, $h=2^\text{nd}$ is always used for the upstream wall.
For computing convective fluxes, the linear and LUST schemes are tested.

The plots in this section will feature the results from all 8 simulations.
To make reading the plots easier, the following line colour and style convention is followed.
Solid lines are used for the results obtained using the linear scheme and dashed lines for those obtained using LUST.
Square markers are used for results using the $n/\delta = 20$ grid, and no markers in the case of $n/\delta =15$.
Finally, red-yellow colour tones are reserved for simulations using $h=1^\text{st}$ and blue-green tones for $h = 2^\text{nd}$.

\begin{figure}[!htb]
\centering
\includegraphics[]{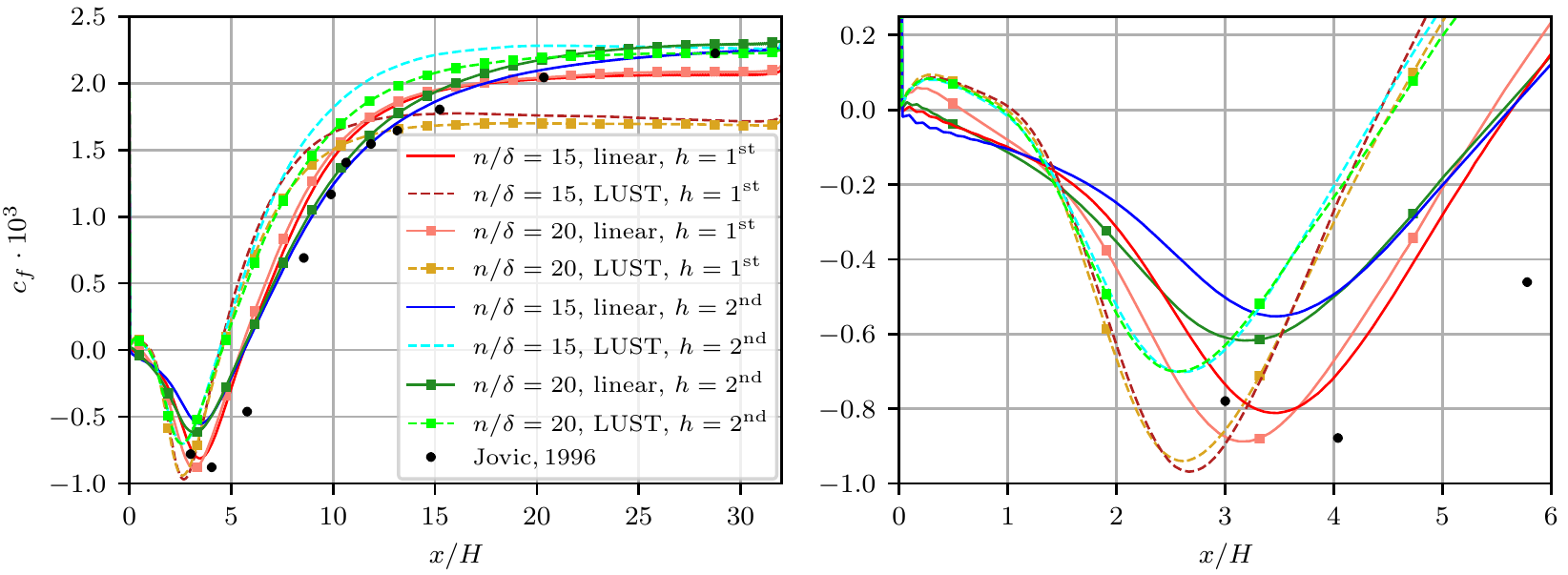}
\caption{Distribution of the skin friction coefficient, $c_f$, over the entire wall downstream of the step (\textit{left}), and over the region of the wall located under the recirculation region (\textit{right}).}
\label{fig:cf}
\end{figure}

\begin{figure}[!htb]
\centering
\includegraphics[]{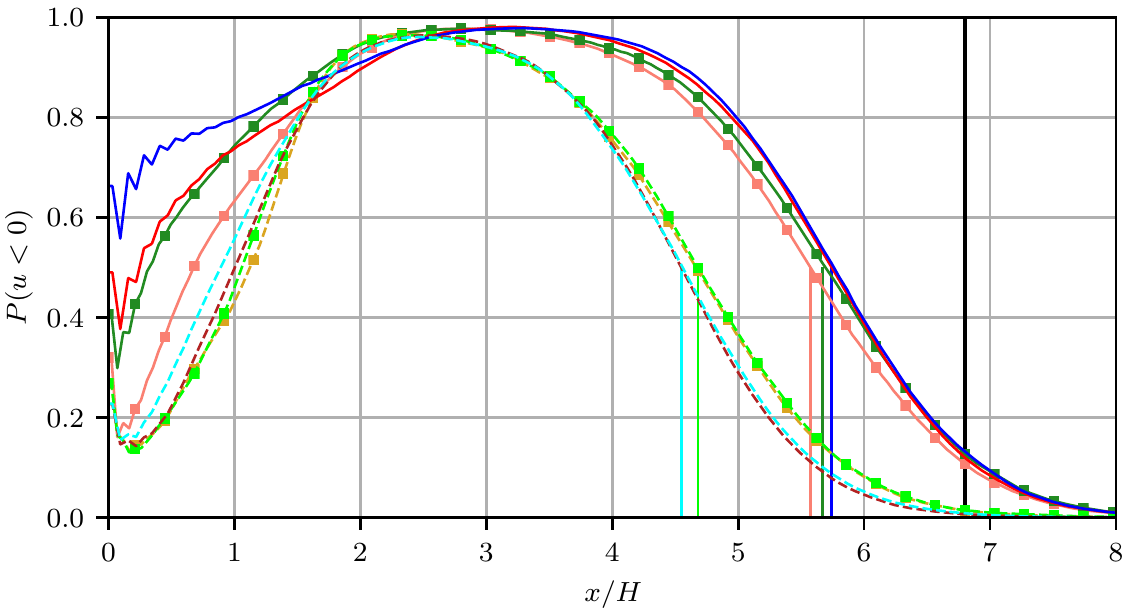}
\caption{The distributions of the probability of back-flow over the flat plate downstream of the step. Line colour and style as in Figure~\ref{fig:cf}.}
\label{fig:backflow}
\end{figure}

Figure~\ref{fig:cf} shows the obtained distributions of the skin friction downstream of the step.
Note that here the abscissa is scaled with $H$.
In the right plot, a zoom into the part of the plate located under the recirculation region is given.
All the simulations using the LUST scheme (dashed lines in the plot) result in positive values of $c_f$ immediately downstream of the step.
When the linear scheme is used (solid lines in the plot), the results vary depending on the choice of $h$ and the density of the grid.
Recall that this growth in $c_f$ is associated with the secondary recirculation bubble.
Examining the distributions of the probability of back-flow in the wall-adjacent cells (see Figure~\ref{fig:backflow}), it is observed that the results in the region occupied by the secondary bubble ($x/H  <1$) are consistent with what is observed for the skin friction.
Only when using the LUST scheme is the presence of the secondary bubble consistently predicted, irrespective of other modelling choices.
The results obtained using the linear scheme, on the other hand, exhibit oscillatory behaviour.
Thus it appears that the numerical oscillations introduced by this scheme on such coarse meshes can be large enough to significantly distort the flow features in this region.
However, using a denser grid (square markers in the plot) does improve the results, and when combined with sampling from the wall adjacent cell both the secondary recirculation bubble and the associated growth in $c_f$ are present.

Attention is now turned to the prediction of the negative values of $c_f$ associated with the main recirculation bubble.
Firstly, it is observed that simulations employing $h=1^\text{st}$ (red-yellow line colours in the plot) result in a larger magnitude of the negative peak, hence leading to better agreement with the reference experimental data.
Recall that in the simulation using $n/\delta = 30$, discussed in the previous section, the grid resolution was not sufficient to resolve the thin boundary layer formed under the recirculation bubble.
It follows that this is also the case for the simulations on coarser grids presented here.
Since above the boundary layer the magnitude of $\mean u$ deceases with $y$ (see Figure~\ref{fig:bfs_u_shear_accurate}), a higher $h$ simply results in lower values of $\mean{\tau_w}$.
Naturally, this result cannot be used as an argument against using $h=2^\text{nd}$ as such, however, unless the grid is adapted to resolve the boundary layer below the recirculation bubble, $h=1^\text{st}$ is a safer choice.  

The size of the recirculation zone can be quantified by the location of the mean reattachment point,~$x_r$.
The latter is here evaluated as the downstream location of $c_f = 0$ and that of $P(u <0) = 0.5$.
The computed values of $x_r/H$ using both methods are presented in Table~\ref{tab:bfs_xr}, with the difference not exceeding $0.15H$.
All simulations produce values that are less than that of the reference experiment, $x_r^0/H = 6.8$.
Evidently, using the LUST scheme systematically leads to the $x_r$-prediction being lower.
It is speculated that this is a result of excessive numerical dissipation introduced by the scheme, leading to damping of turbulent motion in the detached shear layer.
This conclusion is supported by the fact that using a denser grid leads to an increase in the predicted $x_r$.
This is observed both in Figure~\ref{fig:backflow} and also in the result obtained using the $n/\delta = 30$ grid, which is $x_r/H \approx  5.55$, a prediction similar to that obtained using the linear scheme on coarser grids, see Table~\ref{tab:bfs_xr}.

\begin{table}[htb!]
\caption{The location of the mean reattachment point, $x_r/H$, computed as the location of $c_f=0$ and $P(u<0)=0.5$. The value obtained in the reference experiment~\cite{Jovic1996} is $x_r^0/H = 6.8$.}
\centering
\label{tab:bfs_xr}
\begin{tabular}{lcccc}
         & \multicolumn{2}{c}{\textbf{Linear}} & \multicolumn{2}{c}{\textbf{LUST}} \\
$\bm{n/\delta}$, $\bm{h}$ & $c_f=0$                & $P(u<0)=0.5$               & $c_f=0$               & $P(u<0)=0.5$  \\
15, $1^\text{st}$       & 5.59        & 5.74        & 4.43                         & 4.55                        \\
15, $2^\text{nd}$       & 5.59        & 5.76        & 4.41                         & 4.56                        \\
20, $1^\text{st}$       & 5.46        & 5.57        & 4.54                          & 4.66                     \\
20, $2^\text{nd}$       & 5.57        & 5.69        & 4.56                          & 4.68
\end{tabular}
\end{table}

Finally, the performance of the wall model in the region where $P( u < 0) \approx 0$ is considered, which corresponds to $x/H \gtrapprox 8$.
In Figure~\ref{fig:cf}, it is seen that using $h=2^\text{nd}$ (blue-green lines) results in $c_f$ predictions that are in better agreement with the experimental data.
This is consistent with what was reported for channel flow.
Another result matching the observations made for channel flow is the greater sensitivity of $c_f$ to $h$ when the LUST scheme is used and, in particular, that it gets heavily under-predicted when $h = 1^\text{st}$ is employed.

\begin{figure}[!htb]
\centering
\includegraphics[]{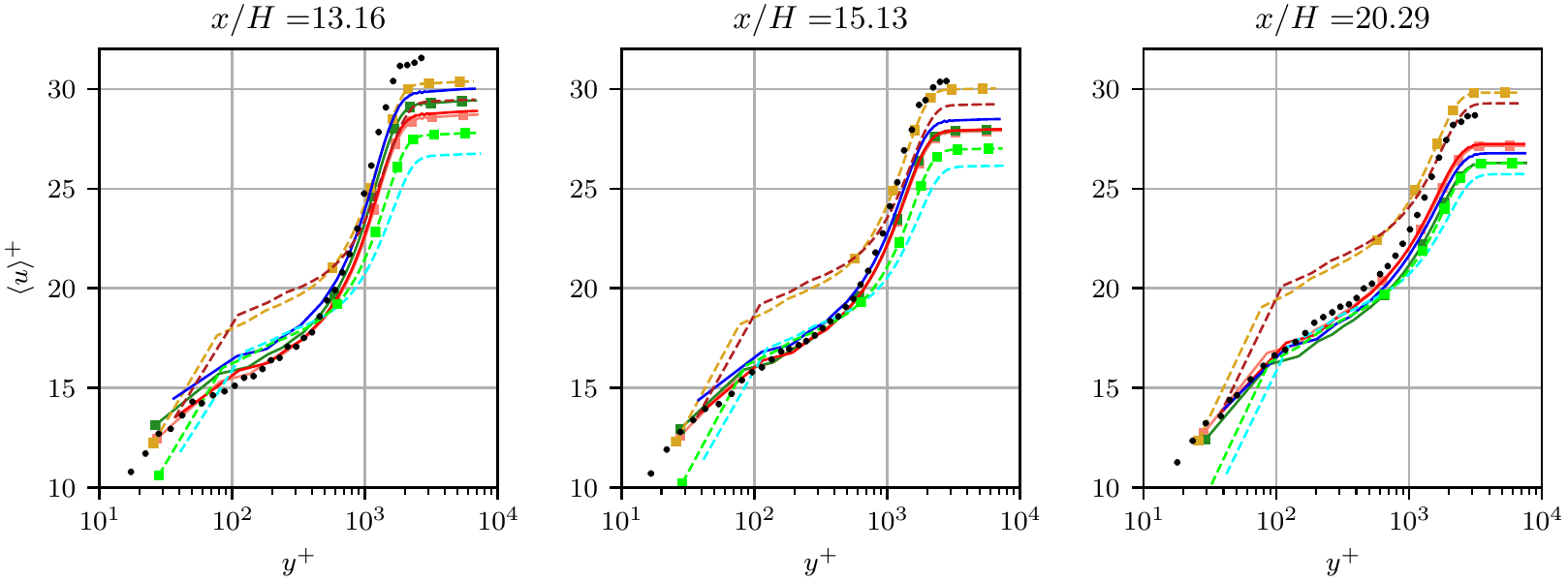}
\caption{Inner-scaled mean streamwise velocity profiles at three stations located downstream of the mean reattachment point. Line colour and style as in Figure~\ref{fig:cf}.}
\label{fig:bfs_u_recovery}
\end{figure}

The latter is also reflected in the inner-scaled mean velocity profiles shown in Figure~\ref{fig:bfs_u_recovery}, see the dashed golden and brown lines (LUST, $h = 1^\text{st}$, $n/\delta = 15$ and 20, respectively).
It is also observed from this figure that when the LUST scheme is used, employing a denser grid results in better agreement with the experimental data.
However, for the linear scheme, no such conclusion can be drawn.
This is also in line with the channel flow results, see Figure~\ref{fig:channel_error_u}. 

In summary, this study has shown that many of the conclusions that were drawn from the channel flow campaign are directly applicable to the more complicated flow over a BFS.
Additionally, it was observed that care must be taken when choosing $h$ in regions where the boundary layer may be heavily under-resolved.
Further, the numerical dissipation introduced by the LUST scheme was shown to result in an under-prediction of $x_r$, when coarse grids are used.

\subsection{Performance of ODE-based models}
\label{sec:bfs_ode}

This section aims to assess the performance of ODE-based models implemented in the library, see Section~\ref{sec:included_wm}.
That is, two treatments of the source term $F$ in equation~\eqref{eq:odetauw} are considered: $F_i = 0$  and $F_i = 1/\rho \partial  p /\partial x_i$, respectively.
Additionally, two models for $\nu_t$ are used, based on equations~\eqref{eq:johnsonking} and~\eqref{eq:duprat}.
The rest of the simulation parameters are as follows.
The $n/\delta =15$ grid is used, along with the linear scheme and sampling from the wall-adjacent cell.
The particular combination of the latter three parameters is not of primary importance for this study since the aim is to focus on the effect of the choice of the wall model.

\begin{figure}[!htb]
\centering
\includegraphics[]{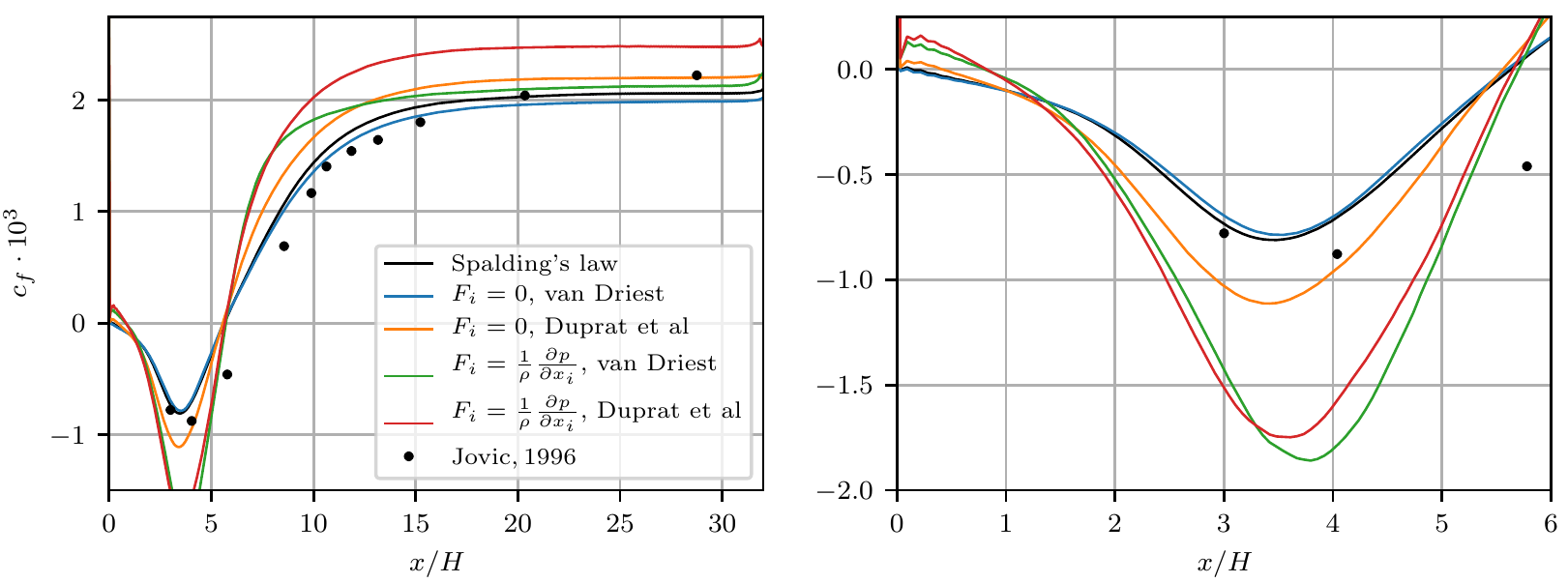}
\caption{The distributions of the skin-friction coefficient computed using ODE-based wall models. The result obtained using Spalding's law is included as reference.}
\label{fig:bfs_cf_ode}
\end{figure}

\begin{figure}[!htb]
\centering
\includegraphics[]{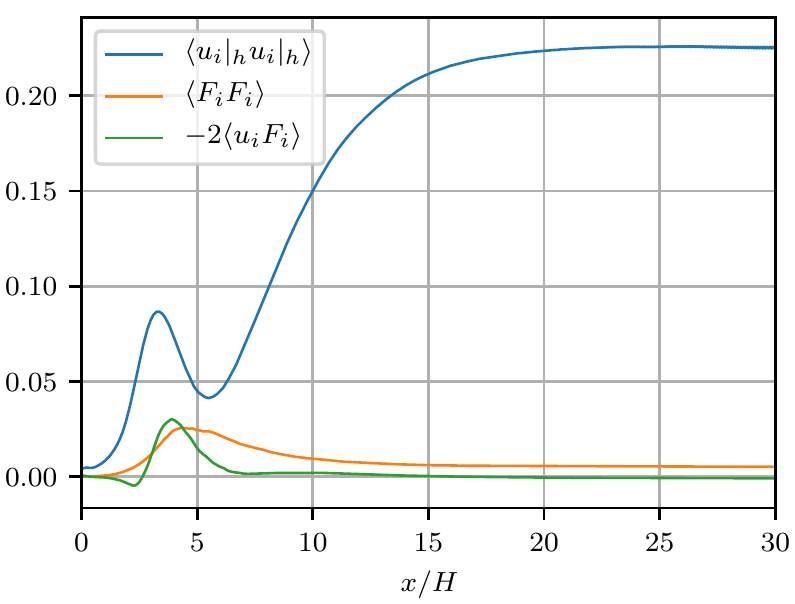}
\caption{The distributions of $\mean{u_i\vert_h u_i\vert_h}$, $\mean{F_i F_i}$ and $-\mean{2u_i\vert_h F_i}$ across the wall downstream of the step.}
\label{fig:bfs_ode_apriori}
\end{figure}

Figure~\ref{fig:bfs_cf_ode} shows the obtained distributions of $c_f$ over the wall downstream of the step.
For reference, the values from a simulation using Spalding's law are also presented.
It is first noted that using $F_i = 0$ and the van Driest-damped mixing length model for $\nu_t$ results in $c_f$ values that are very close to those obtained using Spalding's law.
This is expected since both models assume the velocity to be sampled from an equilibrium zero-pressure-gradient TBL.
Taking into account the pressure gradient leads to an increase in $\mean {\tau_w}$ along the whole wall.
For $x/H < 10$ this is not surprising because the magnitude of the pressure gradient in this region is strong.
However, the large difference observed further downstream is less expected and, therefore, needs further analysis.
Recall that, for the ODE-based wall models, the magnitude of the filtered wall shear stress is obtained as

\begin{align}
    \label{eq:tauodecompact}
& \tau_w =  \left( u_i \vert_h u_i \vert_h + F_iF_iI_1^2  - 2u_i\vert_h F_i I_1  \right)^{1/2} / \left \vert I_2\right \vert,
\end{align}
where $I_1 = \int^h_0 \frac{x_2 }{\nu + \nu_t}\text{d}x_2$, $I_2 = \int^h_0 \frac{1}{\nu + \nu_t}\text{d}x_2$, see~\eqref{eq:odetauw}.
To better understand the behaviour of $\mean {\tau_w}$, the average values of the three quantities in the nominator of~\eqref{eq:tauodecompact} sampled from the LES have been computed during the course of one of the simulations.
The results are shown in Figure~\ref{fig:bfs_ode_apriori}.
One important observation is that the contribution of all three terms is positive, excluding a small region near the step where $-\mean{2u_i\vert_h F_i}$ is negative.
The term $\mean{F_i F_i}$ is not close to zero even for $x/H >10$ and since the mean pressure gradient in that region is negligible, this has to be attributed to the pressure gradient fluctuations.

Note that $\mean{F_i F_i}$ is multiplied by $I_1^2$.
It is possible to get a rough estimate of the mean value of the latter using the $\mean{\tau_w}$-values obtained in the simulation.
Using the conventional $\nu_t$ model based on van Driest damping (equation~\eqref{eq:johnsonking}), $\mean{I_1^2}$ is  predicted to be almost constant for $x/H>10$ and equal to $\approx 13$.
Thus, the contribution of $\mean{F_i F_i}$ is amplified by an order of magnitude, explaining the result in Figure~\ref{fig:bfs_cf_ode}.
Computing the mean values of $I_1^2$, $I_1$ and $I_2$ would be necessary to fully account for the differences in the results from the simulations using the two $\nu_t$ models.
However, since the velocity scale $u_{\tau p}$ used by the model of Duprat et al~\cite{Duprat2011} incorporates the magnitude of the pressure gradient, it is clear that for $x/H > 10$ the discrepancy must be due to pressure gradient fluctuations as well.

Unfortunately, the ODE-based models taking into account only the pressure gradient failed to improve on the results based on simpler algebraic approaches.
However, it should be noted that the flow over a BFS is perhaps not the best case for testing and applying these models.
The pressure gradient is strong directly prior to separation but since the separation point is fixed accounting for it does not significantly affect the results downstream.
After the step, the pressure gradient is  strong in the recirculation zone and a short region downstream of it.
Here, the boundary layer is extremely thin and its physical properties vary significantly from what the considered ODE models were designed to model.
Further downstream, the performance can perhaps be improved by applying a time-filter to the values sampled from the LES, as done in e.g.~\cite{Yang2015, Yang2017}.
This will reduce the effect of the pressure gradient fluctuations on $\mean{\tau_w}$.
However, an improvement upon results given by equilibrium models can hardly be expected.

\section{Conclusions}
\label{sec:conclusions}

This article presents a new open-source library for WMLES, implementing a set of wall-stress models based on both algebraic and ordinary differential equations.
While the majority of the implemented models have already been proposed in the literature, the extension of the integrated formulation of algebraic models to sampling from cells other than the wall-adjacent one is a novelty.
The main advantage of the developed code, besides for its availability for public use, is that it is based on~\ttt{OpenFOAM} technology and thus directly applicable to simulations of both industrial and academic flow cases.
The design of the library is flexible and extendible, which facilitates both exploring the effects of different wall model parameters on the predictive accuracy as well as testing novel wall modelling approaches.

To demonstrate the capabilities of the library, it has been applied to WMLES of fully-developed turbulent channel flow and the flow over a BFS.
For both flows, extensive simulation campaigns have been performed, analysing the effect of mesh resolution, distance to the sampling point of the wall model, and the employed interpolation scheme for the convective fluxes, see Sections~\ref{sec:channel_parameter_study} and~\ref{sec:bfs_parameter_study}.
The choice of the latter was found to be the most influential as it significantly affects not only the predictions of $\mean{\tau_w}$ but also the statistical moments of the velocity field.
Generally, better results for the above quantities were achieved using the LUST scheme.
LUST also led to results consistently improving with mesh refinement and absence of contamination of the flow features with numerical oscillations.
Also, in all the considered cases, sampling from the wall-adjacent cell led to a deterioration of the accuracy of $\mean{ \oltau_w}$ predictions, in line with previously reported results~\cite{Kawai2012, Lee2013, Frere2017}.

Based on the above findings, the recommended choices for the three considered modelling parameters are the LUST scheme, grid density of $n_0 = 27\,000$ cells per $\delta^3$-cube, and $h=2^\text{nd}$.
For channel flow, this combination leads to a $1.25\%$ error in $\mean{u_\tau}$ with respect to DNS data~\cite{Lee2015} and a corresponding error of less than $1\%$ in $\mean{u}/U_b$ in the core of the channel ($y/\delta > 0.2$).
For the flow over a BFS, the errors could not be quantified in the same manner, but the observed agreement in $c_f$ and $\mean u /U_0$ with the experimental data~\cite{Jovic1996} is very good.

Regarding wall modelling, it was shown that simple algebraic wall models are capable of accurately predicting the wall shear stress even when the state of the TBL is far from what the underlying law of the wall assumes it to be.
Employing the integrated form of a given algebraic model seems to lead to improved performance when the sampling of wall model input is done from the wall-adjacent cells, corroborating~\cite{Temmerman2003}.
However, the practice of using those cells for this purpose appears suboptimal, the fundamental reason being that given in~\cite{Kawai2012}, i.e.~associated inaccuracy of the input velocity signal due to numerical errors.
Finally, using ODE-based models attempting to account for the effect of the pressure gradient has, unfortunately, led to degradation in the accuracy of $\mean{\tau_w}$-predictions.
Whether this is an indication that considering only part of the terms composing the right-hand-side of the TBLE equations~\eqref{eq:tble} is incorrect (as advocated for in~\cite{Larsson2016}) or an artefact of excessive fluctuations being fed into the model (see the discussion in Section~\ref{sec:bfs_ode}) requires further study.

Several directions of future work can be identified.
In terms of library development, one is accommodating ODE-based models where the source term $F$ is dependent on the wall-normal coordinate.
Another is developing support for wall modelling approaches that utilise input from more than one sampling point, see~\cite{Rezaeiravesh2018}.
Further validation and improvement of the above-given guidelines for the choices of WMLES modelling parameters is also important.
In particular, considering SGS models other than WALE is necessary.
A more extensive examination of the performance of wall models incorporating the pressure gradient is also needed.
Suitable test-cases are flows exhibiting separation from a curved surface, for example, flow over periodic hills~\cite{Temmerman2003, Frohlich2005, Duprat2011}.

\section{Acknowledgements}
The computations were performed on resources provided by the Swedish National Infrastructure for Computing (SNIC) at PDC Centre for High Performance Computing (PDC-HPC).
The work was supported by Grant No 621-2012-3721 from the Swedish
Research Council.


%
\bibliographystyle{plain}
\bibliography{../../library}

\begin{thebibliography}{10}

\bibitem{Bae2018}
H.~J. Bae, A.~Lozano-Dur{\'{a}}n, S.~T. Bose, and P.~Moin.
\newblock {Turbulence intensities in large-eddy simulation of wall-bounded
  flows}.
\newblock {\em Physical Review Fluids}, 3:014610, 2018.

\bibitem{Balaras1996}
E.~Balaras, C.~Benocci, and U.~Piomelli.
\newblock {Two-layer approximate boundary conditions for large-eddy
  simulations}.
\newblock {\em AIAA Journal}, 34(6):1111--1119, 1996.

\bibitem{Bentaleb2012}
Y.~Bentaleb, S.~Lardeau, and M.~A. Leschziner.
\newblock {Large-eddy simulation of turbulent boundary layer separation from a
  rounded step}.
\newblock {\em Journal of Turbulence}, 13(4), jan 2012.

\bibitem{Bose2014}
S.~T. Bose and P.~Moin.
\newblock {A dynamic slip boundary condition for wall-modeled large-eddy
  simulation}.
\newblock {\em Physics of Fluids}, 26(1):015104, 2014.

\bibitem{Bose2018a}
S.~T. Bose and G.~I. Park.
\newblock {Wall-modeled large-eddy simulation for complex turbulent flows}.
\newblock {\em Annual Review of Fluid Mechanics}, 50(1):535--561, 2018.

\bibitem{Breuer2009}
M.~Breuer, N.~Peller, Ch. Rapp, and M.~Manhart.
\newblock {Flow over periodic hills - Numerical and experimental study in a
  wide range of Reynolds numbers}.
\newblock {\em Computers and Fluids}, 38(2):433--457, 2009.

\bibitem{Cabot1995}
W.~Cabot.
\newblock {Large-eddy simulations with wall models}.
\newblock {\em Annual Research Briefs, Center for Turbulence Research, Stanford
  University}, pages 41--50, 1995.

\bibitem{Cabot1999}
W.~Cabot and P.~Moin.
\newblock {Approximate wall boundary conditions in the large-eddy simulation of
  high Reynolds number flow}.
\newblock {\em Flow, Turbulence and Combustion}, 63:269--291, 1999.

\bibitem{Chapman1979}
D.~R. Chapman.
\newblock {Computational aerodynamics development and outlook}.
\newblock {\em AIAA Journal}, 17(12):1293--1313, 1979.

\bibitem{Choi2012}
H.~Choi and P.~Moin.
\newblock {Grid-point requirements for large eddy simulation: Chapman's
  estimates revisited}.
\newblock {\em Physics of Fluids}, 24(1):011702, 2012.

\bibitem{DeVilliers2006}
E.~{De Villiers}.
\newblock {\em {The Potential of Large Eddy Simulation for the Modeling of Wall
  Bounded Flows}}.
\newblock PhD thesis, Imperial College of Science, Technology and Medicine,
  2006.

\bibitem{Duprat2011}
C.~Duprat, G.~Balarac, O.~Me{\'e}tais, P.~M. Congedo, and O.~Brugi{\`e}re.
\newblock {A wall-layer model for large-eddy simulations of turbulent flows
  with/out pressure gradient}.
\newblock {\em Physics of Fluids}, 23(1):015101, 2011.

\bibitem{Ferziger2002}
J.~H. Ferziger and M.~Peric.
\newblock {\em {Computational Methods for Fluid Dynamics}}.
\newblock Springer-Verlag, 2002.

\bibitem{Frere2017}
A.~Fr{\`{e}}re, C.~C. de~Wiart, K.~Hillewaert, P.~Chatelain, and
  G.~Winckelmans.
\newblock {Application of wall-models to discontinuous Galerkin LES}.
\newblock {\em Physics of Fluids}, 29:085111, 2017.

\bibitem{Frohlich2005}
J.~Fr{\"{o}}hlich, C.~P. Mellen, W.~Rodi, L.~Temmerman, and M.~A. Leschziner.
\newblock {Highly resolved large-eddy simulation of separated flow in a channel
  with streamwise periodic constrictions}.
\newblock {\em Journal of Fluid Mechanics}, 526(2005):19--66, 2005.

\bibitem{Grotzbach1987}
G.~Gr{\"{o}}tzbach.
\newblock {Direct numerical and large eddy simulation of turbulent channel
  flows}.
\newblock In {\em Encyclopedia of Fluid Mechanics}, volume~6, pages 1337--1391.
  1987.

\bibitem{Issa1986}
R.~I. Issa.
\newblock {Solution of the implicitly discretised fluid flow equations by
  operator-splitting}.
\newblock {\em Journal of Computational Physics}, 62(1):40--65, 1986.

\bibitem{Jasak1996}
H.~Jasak.
\newblock {\em {Error Analysis and Estimation for the Finite Volume Method with
  Applications to Fluid Flows}}.
\newblock PhD thesis, Imperial College of Science, Technology and Medicine,
  1996.

\bibitem{Jovic1996}
S.~Jovic.
\newblock {An experimental study of a separated/reattached flow behind a
  backward-facing step. {$Re_h = 37\;000$}}.
\newblock Technical report, NASA Ames Research Center, 1996.

\bibitem{Kawai2012}
S.~Kawai and J.~Larsson.
\newblock {Wall-modeling in large eddy simulation: Length scales, grid
  resolution, and accuracy}.
\newblock {\em Physics of Fluids}, 24(1):015105, 2012.

\bibitem{Kawai2013}
S.~Kawai and J.~Larsson.
\newblock {Dynamic non-equilibrium wall-modeling for large eddy simulation at
  high Reynolds numbers}.
\newblock {\em Physics of Fluids}, 25(1):015105, jan 2013.

\bibitem{Larsson2016}
J.~Larsson, S.~Kawai, J.~Bodart, and I.~Bermejo-Moreno.
\newblock {Large eddy simulation with modeled wall-stress: recent progress and
  future directions}.
\newblock {\em Mechanical Engineering Reviews}, 3(1):1--23, 2016.

\bibitem{Lee2013}
J.~Lee, M.~Cho, and H.~Choi.
\newblock {Large eddy simulations of turbulent channel and boundary layer flows
  at high Reynolds number with mean wall shear stress boundary condition}.
\newblock {\em Physics of Fluids}, 25:110808, 2013.

\bibitem{Lee2015}
M.~Lee and R.~D. Moser.
\newblock {Direct numerical simulation of turbulent channel flow up to
  $\mbox{Re}_\tau \approx 5200$}.
\newblock {\em Journal of Fluid Mechanics}, 774:395--415, 2015.

\bibitem{Liefvendahl2017b}
M.~Liefvendahl and C.~Fureby.
\newblock {Grid requirements for LES of ship hydrodynamics in model and full
  scale}.
\newblock {\em Ocean Engineering}, 143:259--268, 2017.

\bibitem{Liefvendahl2018}
M.~Liefvendahl and M.~Johansson.
\newblock {Wall-modeled LES for ship hydrodynamics in model scale}.
\newblock In {\em 32nd Symposium on Naval Hydrodynamics}, Hamburg, Germany,
  2018.

\bibitem{Liefvendahl2017}
M.~Liefvendahl, T.~Mukha, and S.~Rezaeiravesh.
\newblock {Formulation of a wall model for LES in a collocated finite-volume
  framework}.
\newblock Technical Report 2017-001, Uppsala University, Department of
  Information Technology, 2017.

\bibitem{Manhart2008}
M.~Manhart, N.~Peller, and C.~Brun.
\newblock {Near-wall scaling for turbulent boundary layers with adverse
  pressure gradient : A priori tests on DNS of channel flow with periodic hill
  constrictions and DNS of separating boundary layer}.
\newblock {\em Theoretical and Computational Fluid Dynamics}, 22(3-4):243--260,
  2008.

\bibitem{Martinez2015}
J.~Mart{\'{i}}nez, F.~Piscaglia, A.~Montorfano, A.~Onorati, and S.~M. Aithal.
\newblock {Influence of spatial discretization schemes on accuracy of explicit
  LES: Canonical problems to engine-like geometries}.
\newblock {\em Computers and Fluids}, 117:62--78, 2015.

\bibitem{Mukha2018a}
T.~Mukha, M.~Johansson, and M.~Liefvendahl.
\newblock {Effect of wall-stress model and mesh-cell topology on the predictive
  accuracy of LES of turbulent boundary layer flows}.
\newblock In {\em 7th European Conference on Computational Fluid Dynamics},
  Glasgow, UK, 2018.

\bibitem{Mukha2017}
T.~Mukha and M.~Liefvendahl.
\newblock {The generation of turbulent inflow boundary conditions using
  precursor channel flow simulations}.
\newblock {\em Computers and Fluids}, 156:21--33, 2017.

\bibitem{Mukha2017a}
T.~Mukha, S.~Rezaeiravesh, and M.~Liefvendahl.
\newblock {An OpenFOAM library for wall-modelled large-eddy simulation}.
\newblock In {\em 12th OpenFOAM Workshop}, Exeter, UK, 2017.

\bibitem{Mukha2018b}
T.~Mukha, S.~Rezaeiravesh, and M.~Liefvendahl.
\newblock {Wall-modelled large-eddy simulation of the flow over a
  backward-facing step}.
\newblock In {\em 13th OpenFOAM Workshop}, Shanghai, China, 2018.

\bibitem{Nicoud1999a}
F.~Nicoud and F.~Ducros.
\newblock {Subgrid-scale stress modelling based on the square of the velocity
  gradient tensor}.
\newblock {\em Flow, Turbulence and Combustion}, 62(3):183--200, 1999.

\bibitem{Nikitin2000}
N.~V. Nikitin, F.~Nicoud, B.~Wasistho, K.~D. Squires, and P.~R. Spalart.
\newblock {An approach to wall modeling in large-eddy simulations}.
\newblock {\em Physics of Fluids}, 12(7):1629--1632, 2000.

\bibitem{Park2014}
G.~I. Park and P.~Moin.
\newblock {An improved dynamic non-equilibrium wall-model for large eddy
  simulation}.
\newblock {\em Physics of Fluids}, 26:015108, jan 2014.

\bibitem{Park2016}
G.~I. Park and P.~Moin.
\newblock {Numerical aspects and implementation of a two-layer zonal wall model
  for LES of compressible turbulent flows on unstructured meshes}.
\newblock {\em Journal of Computational Physics}, 305:589--603, 2016.

\bibitem{Piomelli2008}
U.~Piomelli.
\newblock {Wall-layer models for large-eddy simulations}.
\newblock {\em Progress in Aerospace Sciences}, 44(6):437--446, 2008.

\bibitem{Piomelli2002}
U.~Piomelli and E.~Balaras.
\newblock {Wall-layer models for large-eddy simulations}.
\newblock {\em Annual Review of Fluid Mechanics}, 34:349--374, 2002.

\bibitem{Pope2000}
S.~B. Pope.
\newblock {Turbulent Flows}.
\newblock {\em Cambridge University Press}, 2000.

\bibitem{Reichardt1951}
H.~Reichardt.
\newblock {Vollst\"{a}ndige Darstellung der turbulenten
  Geschwindigkeitsverteilung in glatten Leitungen}.
\newblock {\em Zeitschrift f\"{u}r Angewandte Mathematik und Mechanik},
  31(7):208--219, 1951.

\bibitem{Rezaeiravesh2017}
S.~Rezaeiravesh and M.~Liefvendahl.
\newblock {Grid construction strategies for wall-resolving large eddy
  simulation and estimates of the resulting number of grid points}.
\newblock Technical report, Uppsala University, Department of Information
  Technology, 2017.

\bibitem{Rezaeiravesh2016}
S.~Rezaeiravesh, M.~Liefvendahl, and C.~Fureby.
\newblock {On grid resolution requirements for LES of wall-bounded flows}.
\newblock In {\em ECCOMAS Congress 2016}, Crete, Greece, 2016.

\bibitem{Rezaeiravesh2018}
S.~Rezaeiravesh, T.~Mukha, and M.~Liefvendahl.
\newblock {a-Priori study of wall modeling in large eddy simulation}.
\newblock In {\em 7th European Conference on Computational Fluid Dynamics},
  Glasgow, UK, 2018.

\bibitem{Sagaut2005}
P.~Sagaut.
\newblock {\em {Large Eddy Simulation for Incompressible Flows: An
  Introduction}}.
\newblock Springer-Verlag, 2005.

\bibitem{Schumann1975}
U.~Schumann.
\newblock {Subgrid scale model for finite difference simulations of turbulent
  flows in plane channels and annuli}.
\newblock {\em Journal of Computational Physics}, 18(4):376--404, 1975.

\bibitem{Song2000}
S.~Song, D.~B. DeGraaff, and J.~K. Eaton.
\newblock {Experimental study of a separating, reattaching, and redeveloping
  flow over a smoothly contoured ramp}.
\newblock {\em International Journal of Heat and Fluid Flow}, 21:512--519,
  2000.

\bibitem{Spalart1997}
P.~R. Spalart, W.~H. Jou, M.~Kh. Strelets, and S.~R. Allmaras.
\newblock {Comments on the feasibility of LES for wings and on a hybrid
  {RANS/LES} approach}.
\newblock In {\em Advances in DNS/LES}, volume~1, 1997.

\bibitem{Spalding1961}
D.~B. Spalding.
\newblock {A single formula for the ``law of the wall''}.
\newblock {\em Journal of Applied Mechanics}, 28(3):455--458, 1961.

\bibitem{Temmerman2003}
L.~Temmerman, M.~A. Leschziner, C.~P. Mellen, and J.~Fr{\"{o}}hlich.
\newblock {Investigation of wall-function approximations and subgrid-scale
  models in large eddy simulation of separated flow in a channel with
  streamwise periodic constrictions}.
\newblock {\em International Journal of Heat and Fluid Flow}, 24(2):157--180,
  2003.

\bibitem{vanDriest1956}
E.~R. van Driest.
\newblock {On turbulent flow near a wall}.
\newblock {\em Journal of the Aeronautical Sciences}, 23(11):1007--1011, 1956.

\bibitem{Wang2002}
M.~Wang and P.~Moin.
\newblock {Dynamic wall modeling for large-eddy simulation of complex turbulent
  flows}.
\newblock {\em Physics of Fluids}, 14:2043, 2002.

\bibitem{Weller2012}
H.~Weller.
\newblock {Controlling the computational modes of the arbitrarily structured C
  grid}.
\newblock {\em Monthly Weather Review}, 140(10):3220--3234, 2012.

\bibitem{Weller1998}
H.~G. Weller, G.~Tabor, H.~Jasak, and C.~Fureby.
\newblock {A tensorial approach to computational continuum mechanics using
  object-oriented techniques}.
\newblock {\em Computers in Physics}, 12(6):620--631, 1998.

\bibitem{Werner1991}
H.~Werner and H.~Wengle.
\newblock {Large-eddy simulation of turbulent flow over and around a cube in a
  plate channel}.
\newblock In {\em Turbulent Shear Flows 8}, pages 155--168. Springer-Verlag,
  1991.

\bibitem{Wu2013}
P.~Wu and J.~Meyers.
\newblock {A constraint for the subgrid-scale stresses in the logarithmic
  region of high Reynolds number turbulent boundary layers: A solution to the
  log-layer mismatch problem}.
\newblock {\em Physics of Fluids}, 25:015104, 2013.

\bibitem{Yang2017}
X.~I.~A. Yang, G.~I. Park, and P.~Moin.
\newblock {Log-layer mismatch and modeling of the fluctuating wall stress in
  wall-modeled large-eddy simulations}.
\newblock {\em Physical Review Fluids}, 2(10):1--13, 2017.

\bibitem{Yang2015}
X.~I.~A. Yang, J.~Sadique, R.~Mittal, and C.~Meneveau.
\newblock {Integral wall model for large eddy simulations of wall-bounded
  turbulent flows}.
\newblock {\em Physics of Fluids}, 27:025112, 2015.

\end{thebibliography}


\begin{thebibliography}{0}
\bibitem{1}J. Larsson, S. Kawai, J. Bodart, and I. Bermejo-Moreno. Large eddy simulation with modeled wall-stress: recent progress and future directions. Mechanical Engineering Reviews, 3(1):1-23, 2016.         
\bibitem{2}J. Slotnick, A. Khodadoust, J. Alonso, D. Darmofal, W. Gropp, E. Lurie, D. Mavriplis.  CFD vision 2030 study: A path to revolutionary computational aerosciences, Tech. rep., NASA, 2014.         
\bibitem{3} S.~T. Bose and G.~I. Park. Wall-modeled large-eddy simulation for complex turbulent flows. Annual Review of Fluid Mechanics, 50(1):535--561, 2018.         
\end{thebibliography}

%

\end{document}